\documentclass[useAMS,usenatbib]{mn2e}
\usepackage{verbatim,epsfig}

\def\nms{\mathsurround=0pt}
\def\oversim#1#2{\lower 2pt\vbox{\baselineskip 0pt \lineskip 1pt
    \ialign{$\nms#1\hfil##\hfil$\crcr#2\crcr\sim\crcr}}}
\def\lesssim{\mathrel{\mathpalette\oversim<}} 
\def\gtrsim{\mathrel{\mathpalette\oversim>}} 

\title[Long-period variables in M15]{Discovery of long-period variable stars in the very-metal-poor globular cluster M15}
\author[I. McDonald et al.]{I. McDonald$^{1,2}$\thanks{E-mail: mcdonald@jb.man.ac.uk}, J.Th. van Loon$^{1}$, A.K. Dupree$^{3}$, M.L. Boyer$^{4}$\\
$^{1}$Astrophysics Group, Lennard-Jones Laboratories, Keele University, Staffordshire, ST5 5BG, UK\\
$^{2}$Jodrell Bank Centre for Astrophysics, Alan Turing Building, Manchester University, M13 9PL, UK\\
$^{3}$Harvard-Smithsonian Center for Astrophysics, 60 Garden Street, Cambridge, MA 02138, USA\\
$^{4}$STScI, 3700 San Martin Drive, Baltimore, MD 21218, USA}
\begin{document}

\date{Accepted 9999 December 31. Received 9999 December 31; in original form 9999 December 31}

\pagerange{\pageref{firstpage}--\pageref{lastpage}} \pubyear{9999}

\maketitle

\label{firstpage}

\begin{abstract}
We present a search for long-period variable (LPV) stars among giant branch stars in M15 which, at [Fe/H] $\sim$ --2.3, is one of the most metal-poor Galactic globular clusters. We use multi-colour optical photometry from the 0.6-m Keele Thornton and 2-m Liverpool Telescopes. Variability of $\delta$V $\sim$ 0.15 mag is detected in K757 and K825 over unusually-long timescales of nearly a year, making them the most metal-poor LPVs found in a Galactic globular cluster. K825 is placed on the long secondary period sequence, identified for metal-rich LPVs, though no primary period is detectable. We discuss this variability in the context of dust production and stellar evolution at low metallicity, using additional spectra from the 6.5-m Magellan (Las Campanas) telescope. A lack of dust production, despite the presence of gaseous mass loss raises questions about the production of dust and the intra-cluster medium of this cluster.
\end{abstract}


\section{Introduction}

The onset of radial pulsation on the red giant branches is important in stellar evolution: it is one of several linked phenomena (including dust production and substantial mass loss) that control the endpoint of stellar evolution and injection of mass into the interstellar medium. While these phenomena are linked, the relative timing of their appearance is poorly understood \citep{MvL07}. In particular, it is debatable whether pulsation can provide enough energy to assist mass loss in these evolved stars (e.g.~\citealt{Bowen88}).

Optical photometric variability in highly-evolved stars is known to be less pronounced in metal-poor systems. \citet{FW98} considered Long-Period Variable stars (LPVs) in both the Galactic Disc and globular clusters in order to prove this dependence, though they only consider stars showing large-amplitude variability as LPVs and do not consider semi-regular variables (SRVs, which are included in the definition of LPVs for the purposes of this work). This raises the question of whether pulsation is capable, or required, to drive mass loss from metal-poor stars.

We describe here the search for LPVs in one of the most metal-poor Galactic globular clusters, M15. This cluster is the only one known to harbour a dusty and/or gaseous interstellar medium and has several infrared-excessive giant stars (\citealt{ESvL+03}; \citealt{vLSEM06}; \citealt{BWvL+06}), giving the strong implication that dusty stellar winds are present in the cluster and, by further implication, pulsation-driven winds.

No LPVs have so-far been found in M15, though several candidates were identified by \citet{MW75} (Table \ref{M15CudworthTable}). These were followed-up by \citet{Welty85}, who could not find photometric variability in three targets (K169, K288 and K709) and retained K757 and K825 as candidate variables (identifiers from \citealt{Kustner21}).

\section{Observations}

\subsection{Liverpool Telescope}

The first dataset for this work comes from the two-metre Liverpool Telescope (LT; \citealt{SSR+04}), situated on La Palma. A total of 38 observation blocks were taken, spanning 462 days, from 2007 April 23 to 2008 July 27. A gap is present between 2007 December 04 and 2008 April 23 when the cluster was viewed in too close proximity to the Sun. Observations were taken roughly every eight days, with the exception of a one-week block of daily observations in August 2007, to build in redundancy against shorter-period variability. Scheduling and weather constraints mean the observations are randomly distributed enough not to form strong aliases in a Fourier spectrum.

Each observation consisted of 3 $\times$ 6\,s $g^\prime$ (477 nm) and 3 $\times$ 5\,s $i^\prime$ (762.5 nm) exposures, thus minimising stochastic effects from cosmic rays, etc. An additional 2\,s image was taken in both filters in case the stars saturated the detector. Preliminary analysis showed this was not necessary and these were not taken in the 2008 observing season. Image calibration was performed automatically before receipt. 

\subsection{Keele Thornton Telescope}

Additional images were taken with Keele's 24-inch (60-cm) Thornton Telescope (KT) sited on Keele University campus, with elevated (205m above sea level) views over the Staffordshire countryside. Although these images have poorer photometric accuracy and astrometric resolution, they cover a longer timebase. The data were taken using the Santa Barbara Instrument Group ST7 CCD camera (at Newtonian focus). Several features were found that have the potential to affect the photometry: namely tracking, coolant issues and random bias and dark current problems. In practice, however, we find that the statistical scatter in photometry from this telescope is smaller than that from the LT.

With these issues in mind, observation blocks were devised of multiple 10-second observations, taken in $V$-, $R$- and $I$-bands (550, 700 and 880 nm, respectively). Observations taken at low altitude have not been corrected for differential reddening (due to its night-to-night variation): some of the scatter in the resultant photometry may be attributable to it. 

In total, 76 observation blocks were taken, spanning from 2006 January 14 to 2008 November 24, though most observations were taken after 2007 August 02. Weather constraints at Keele are significantly worse than those on La Palma, leading to even less-regular sampling.

\begin{figure}
\centerline{\resizebox{\hsize}{!}{\includegraphics[angle=0]{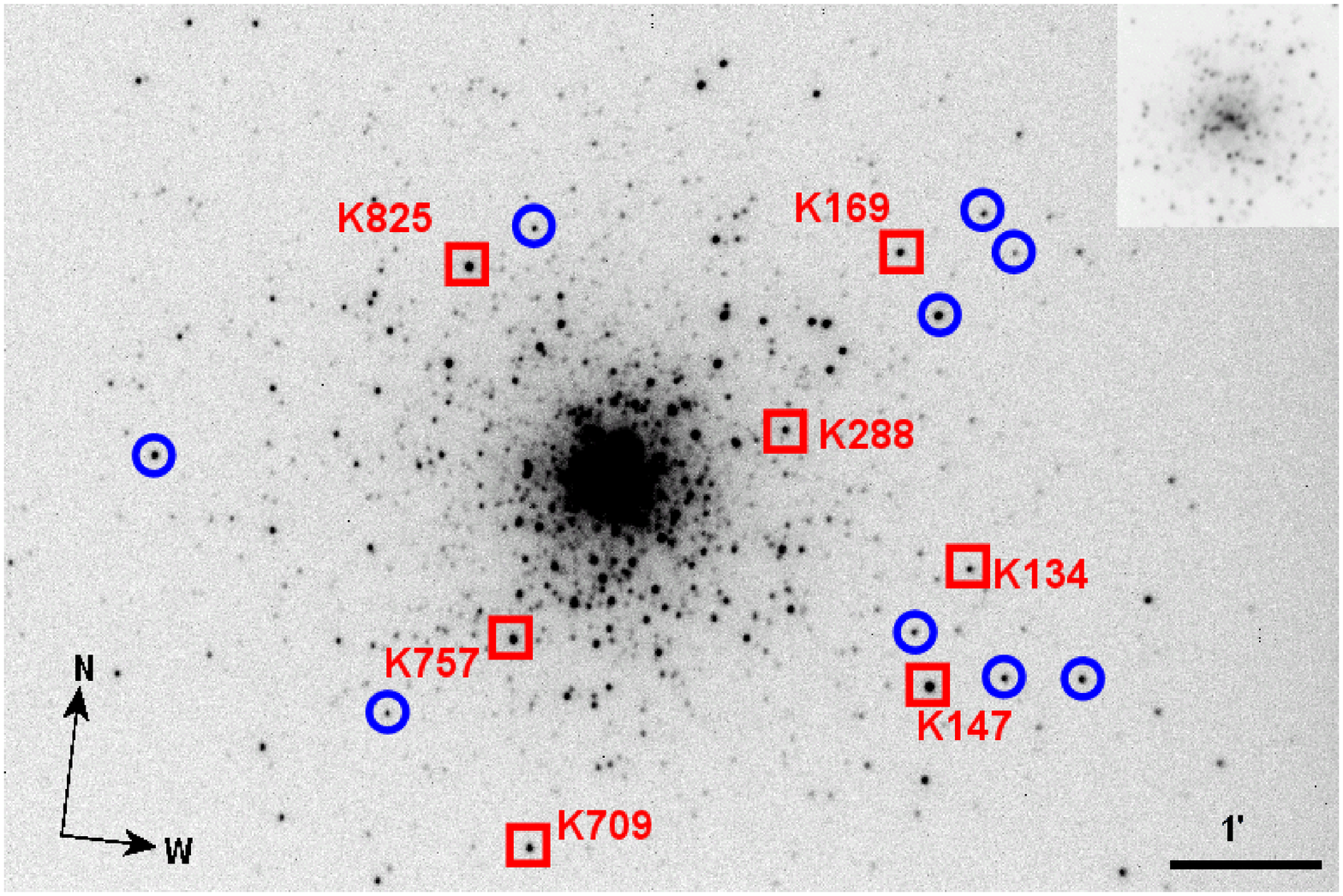}}}
\centerline{\resizebox{\hsize}{!}{\includegraphics[angle=0]{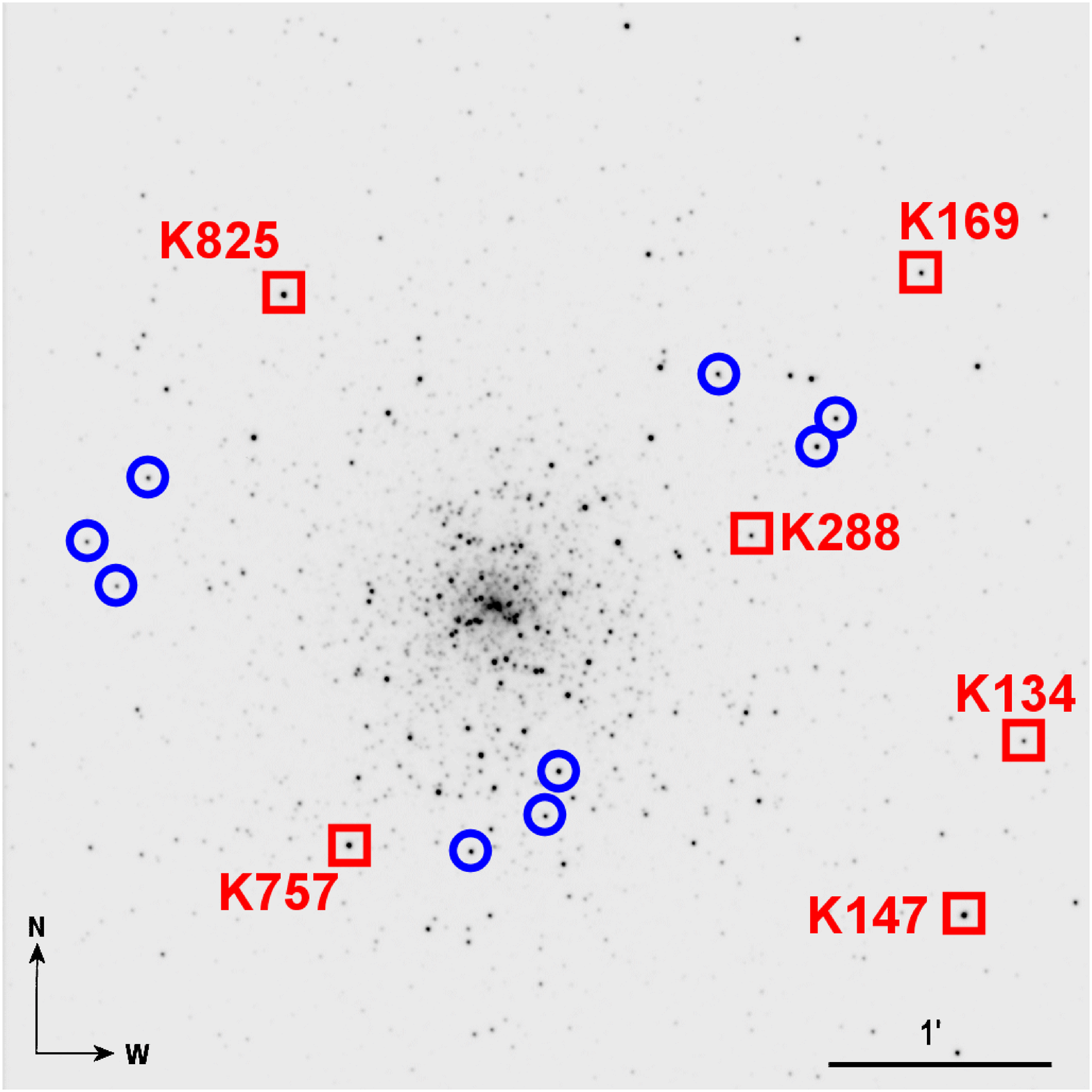}}}
 \caption[Example images of M15 from Keele and the LT]{Negative example images of M15 from the Keele 24-inch Thornton Telescope (top panel; $V$-band image) and the Liverpool Telescope (bottom panel; average of $g^\prime$ and $i^\prime$ images). Comparison stars (blue circles) and candidate variables (red squares) from \protect\cite{MW75} are shown. The inset in the Keele figure shows the heavy crowding in the core. The Liverpool Telescope figure is displayed with a gamma-law scaling to aid viewing.}
 \label{M15ImageFig}
\end{figure}

\section{Results}

\subsection{Candidate selection}

The presence of variability can be determined via difference imaging. Here, an image (\emph{A}) is scaled and convolved to match the point-spread function of a second image (\emph{B}, taken at a different epoch). The resulting image is subtracted from image \emph{B} to create a difference image. Variable stars show up as non-zero flux.

We performed difference imaging on pairs of images from both telescopes, taken at various epochs. Variability was immediately identifiable in K825 and suggested in K757. Of the other stars in the cluster, the only obvious suggestion of variability came from the cluster's core, which is entirely unresolved in our images. We therefore do not consider variability in stars other than those candidate variables listed in \citet{MW75}, which include K757 and K825.

\subsection{Photometric reduction}

\subsubsection{Comparison star selection}
\label{PRCS}

Example images from both telescopes are shown in Fig.\ \ref{M15ImageFig}. Aperture photometry was extracted from these images using the software {\sc AIP4Win} v.\,1.3.5 \citep{BB05}. Comparison stars (Table \ref{ComparisonTable}, Fig.\ \ref{M15ImageFig}) were chosen to be:

\begin{list}{$\bullet$}{}
	\item isolated (resolved from any visible neighbours by at least 2.5$\times$ the full-width half-maximum seeing), so that seeing has minimal effect on photometry due to blended stellar images;
	\item bright ($i^\prime < 14$ mag), to provide good signal-to-noise, but not saturated;
	\item red ($B-V > 0.75$ mag), to avoid problems with differential reddening at high airmass;
	\item invariable, nor variable candidates themselves (we retain K1040 as its known short-period variability of 0.04 mag is well below our sensitivity --- \citealt{Bao-An90}; \citealt{YZQT93}).
\end{list}


\begin{center}
\begin{table*}
\caption[M 15 Literature Candidate Giant Variables]{Candidate variable giant stars in M15 from \cite{MW75} with variability results. Proper motion memberships and colours are from \cite{Cudworth76}. Other references and notes as Table \protect\ref{ComparisonTable}. Note that all periods are very approximate.}
\label{M15CudworthTable}
\begin{tabular}{llllllcr@{}l@{\ \ }c@{\ \ }c@{\ \ }c@{\ \ }c}
    \hline \hline
\multicolumn{4}{l}{Designator}	& \multicolumn{2}{c}{Co-ordinates (J2000)}	
			& Membership	& \multicolumn{2}{c}{v}	
			& $V$ & ($B-V$)
			& Variability	& Period\\
(1)	&	(2)	&	(3)	&	(4)	
			&	\multicolumn{1}{c}{RA}	&	\multicolumn{1}{c}{Dec}
			& probability	& \multicolumn{2}{c}{(km s$^{-1}$)}
			&(mag) & (mag) 
			& found?	& (days)\\
    \hline
K134	&	III-8	& 416	& \ 	& 21\,29\,49.07 & 12\,09\,02.0	& $\sim$0\%	&	\ &\ 		& 14.02	& 0.99	& No &	\ \\
K147	&	III-34& 442	& \ 	& 21\,29\,50.01 & 12\,08\,43.4	& $\sim$0\%		&	+19&$^5$	& 12.62	& 0.75	& Maybe & \ \\
K169	&	II-64	& 212	& \ 	& 21\,29\,50.81 & 12\,11\,30.0	& \ 60\%		&	\ &\ 		& 13.48	& 1.12	& No &	\\
K288	&	II-16	& \ 	& \ 	& 21\,29\,53.78 & 12\,10\,20.1	& \ 90\%	&	--107&.2$^6$	& 13.59	& 1.06	& No &	\ \\
K709	&	IV-58	& \ 	& \ 	& 21\,30\,00.39 & 12\,07\,36.0	& \ 98\%	&	--100&.7$^7$	& 13.52	& 1.07	& Maybe & \ \\
K757	&	IV-38	& \ 	& \ 	& 21\,30\,00.91 & 12\,08\,56.8	& \ 95\%	&	--113&.5$^6$	& 12.58	& 1.36	& Yes & $\sim$250?\\
K825	&	I-12	& \ 	&S4	& 21\,30\,02.23 & 12\,11\,21.5	& \ 67\%	&	--98&.87$^8$	& 12.69	& 1.37	& Yes & 350$\pm$20\\
\hline
\multicolumn{13}{p{0.90\textwidth}}{\small $^1$\cite{Kustner21}; $^2$\cite{Arp55}; $^3$\cite{BBC+83}; $^4$\cite{Sandage70}; heliocentric radial velocities from $^5$\cite{SKLT02}; $^6$\cite{GPW+97}; $^7$\cite{MDS08}; and $^8$\cite{PSK+00}; cf.~M15's velocity of $v_{\rm heliocentric} = -107.9$ km s$^{-1}$ and internal velocity dispersion of $\sim$10 km s$^{-1}$ \citep{GPW+97}. \normalsize} \\
    \hline
\end{tabular}
\end{table*}
\end{center}

\begin{center}
\begin{table}
\caption[Comparison stars for photometric reduction (Keele)]{Comparison stars used in the photometric reduction of images from the Keele Thornton and Liverpool Telescopes. K497, K508 and K589 were not used in the final analysis due to possible intrinsic variability.}
\label{ComparisonTable}
\begin{tabular}{l@{\ }l@{\ \,}l@{\ \ }l@{\ \ \ }l@{\ }l@{\ \ \ }l@{\ \ }l@{\ }}
    \hline \hline
\multicolumn{4}{l}{Designator}	& \multicolumn{2}{l}{Co-ordinates (J2000)}	& $V$	& \llap{($B$} -- $V$)  \\
(1)	& (2)	& (3)	& (4)	&	RA	&	Dec	& (5)	& (5) 	\\
    \hline
    \multicolumn{8}{c}{Comparison stars (Keele)}\\
K87	& III	& 438	& S29	& 21\,29\,45.81\,& 12\,08\,45.1\,& 13.81 & 1.01\\
K112	& II	& 213	& \ 	& 21\,29\,47.90\,& 12\,11\,31.3\,& 15.12 & 0.88$^3$\\
K114	& III	& 437	& \ 	& 21\,29\,47.86\,& 12\,08\,45.0\,& 13.87 & 1.06\\
K129	& II	& 187	& \ 	& 21\,29\,48.61\,& 12\,11\,45.6\,& 14.27 & 0.96\\
K144	& II-75	& 248	& \ 	& 21\,29\,49.78\,& 12\,11\,05.6\,& 13.00 & 1.25\\
K158	& III-33& 414	& \ 	& 21\,29\,50.27\,& 12\,09\,02.5\,& 14.13 & 0.78\\
K731	& I-63	& \ 	& \ 	& 21\,30\,00.52\,& 12\,11\,36.9\,& 14.24 & 0.96\\
K912	& IV-48	& 457	& \ 	& 21\,30\,04.20\,& 12\,08\,27.4\,& 14.40 & 0.91\\
K1040	& I	& 319	& S6	& 21\,30\,10.49\,& 12\,10\,06.2\,& 13.40 & 1.18\\
    \hline
    \multicolumn{8}{c}{Comparison stars (Liverpool Telescope)}\\
 K224	& II-30 & \ 	& \ 	& 21\,29\,52.30 & 12\,10\,51.3 & 14.51	& 1.07$^7$ \\
 K238	& II-29	& \ 	& \ 	& 21\,29\,52.63 & 12\,10\,43.8 & 14.46	& 1.17$^7$ \\
 K319	& II	& \ 	& \ 	& 21\,29\,54.42 & 12\,11\,02.4 & \     & \  \\
 K497	& III 	& \ 	& \ 	& 21\,29\,57.17 & 12\,09\,17.3 & 14.42	& 1.29$^7$ \\
 K508	& III 	& \ 	& \ 	& 21\,29\,57.49 & 12\,09\,06.1 & \     & \  \\
 K589	& IV	& \ 	& \ 	& 21\,29\,58.71 & 12\,08\,55.8 & \     & \  \\
 K928	& I-43	& \ 	& \ 	& 21\,30\,04.61 & 12\,10\,32.6 & 13.83 & 1.03 \\
 K943	& I-38 	& \ 	& \ 	& 21\,30\,05.14 & 12\,10\,04.1 & 14.24 & 0.94 \\
 K961	& I-41	& \ 	& \ 	& 21\,30\,05.69 & 12\,10\,15.7 & 14.09 & 1.03 \\
    \hline
\multicolumn{8}{p{0.45\textwidth}}{\small $^1$\citet{Kustner21}; $^2$\citet{Arp55} (incomplete designations denote the quadrant of Arp's map containing the star if the star is without a designation); $^3$\citet{BBC+83}; $^4$\citet{Sandage70}; $^5$\citet{Cudworth76}; $^7$from \emph{photographic and photovisual} magnitudes from \citet{Brown51}. \normalsize} \\
	 \hline
\end{tabular}
\end{table}
\end{center}

\subsubsection{Keele Thornton Telescope}
\label{PRTT}

The obvious problem is that bright, red stars tend to be other cluster giants, which may also vary. Our comparison stars show no intrinsic variation, though a systematic offset in several of the comparison stars (K129, K144 and K912) appears depending on the orientation of the telescope. The reason for this is unclear, though seeing and airmass may play a r\^ole: the telescope must be inverted at this declination to observe at higher airmasses in the east, where light pollution is greatest. There may also be a weak correlation between calculated differential magnitude and seeing quality. Due to concerns that this offset is also present in the candidate stars, we have discounted the frames taken when the telescope was inverted from further analysis.

Images were calibrated and stacked using {\sc AIP4Win} to increase signal-to-noise. Though not a lossless procedure, tests on individual and stacked images shows that the likely effect on the results is $<$1 milli-mag (mmag). A standard photometric reduction was performed: i.e.\ bias and dark frame subtraction, and flatfield division. Aperture photometry was performed using {\sc AIP4Win}. Photometric errors (not provided by {\sc AIP4Win}) were calculated manually by adding in quadrature the relative Poissonian errors in the photometric signal and calibration frames.

The absence of \emph{long-}period variability of the comparison stars K144 (Arp II-75) and K1040 is interesting in itself, as these stars are identified as being among the brightest, reddest variables in the cluster by \citet{FPC83}. (Although K1040 (S6) is a known short-period variable (\S\ref{PRCS}), the 4.3-hour period is too short and the $\delta$V = 0.04 mag amplitude too small to affect our data.)

An identical photometric reduction was done on the candidate variable stars. The flux of each of the candidate variables was divided by the sum of the fluxes of the comparison stars to yield a differential magnitude.

\subsubsection{Liverpool Telescope}

\begin{figure*}
\centerline{\resizebox{\hsize}{!}{\includegraphics[angle=270]{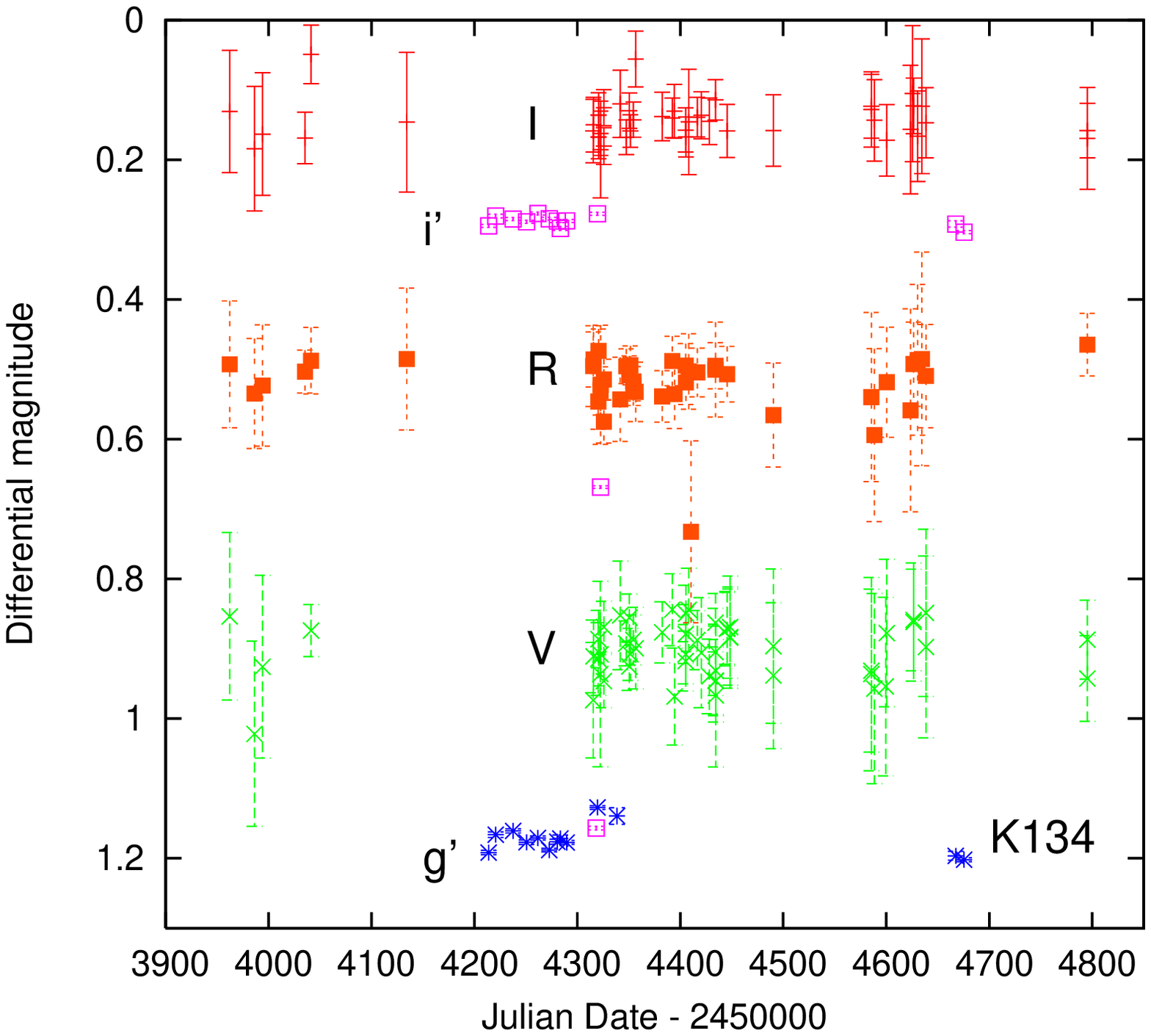}\includegraphics[angle=270]{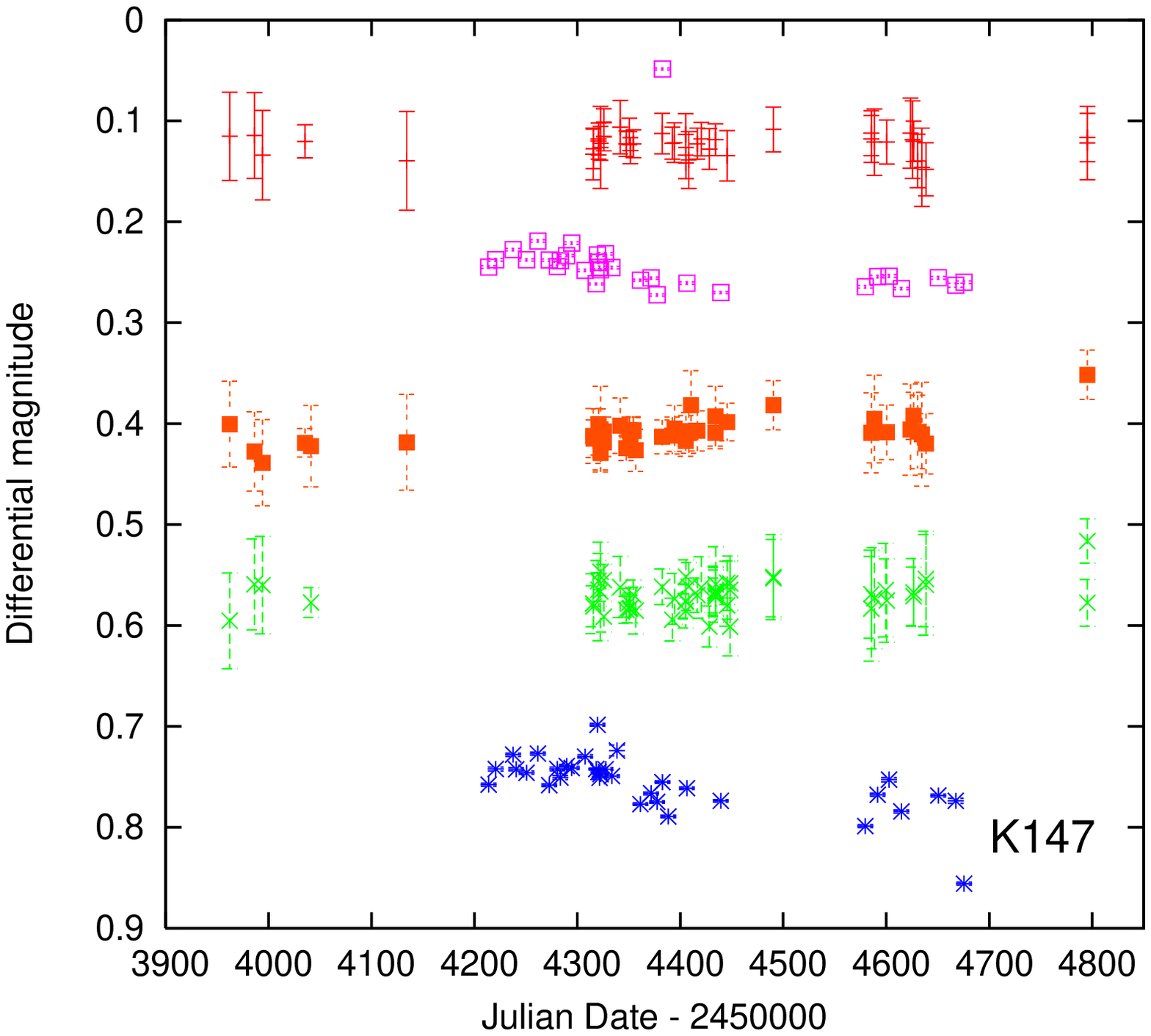}\includegraphics[angle=270]{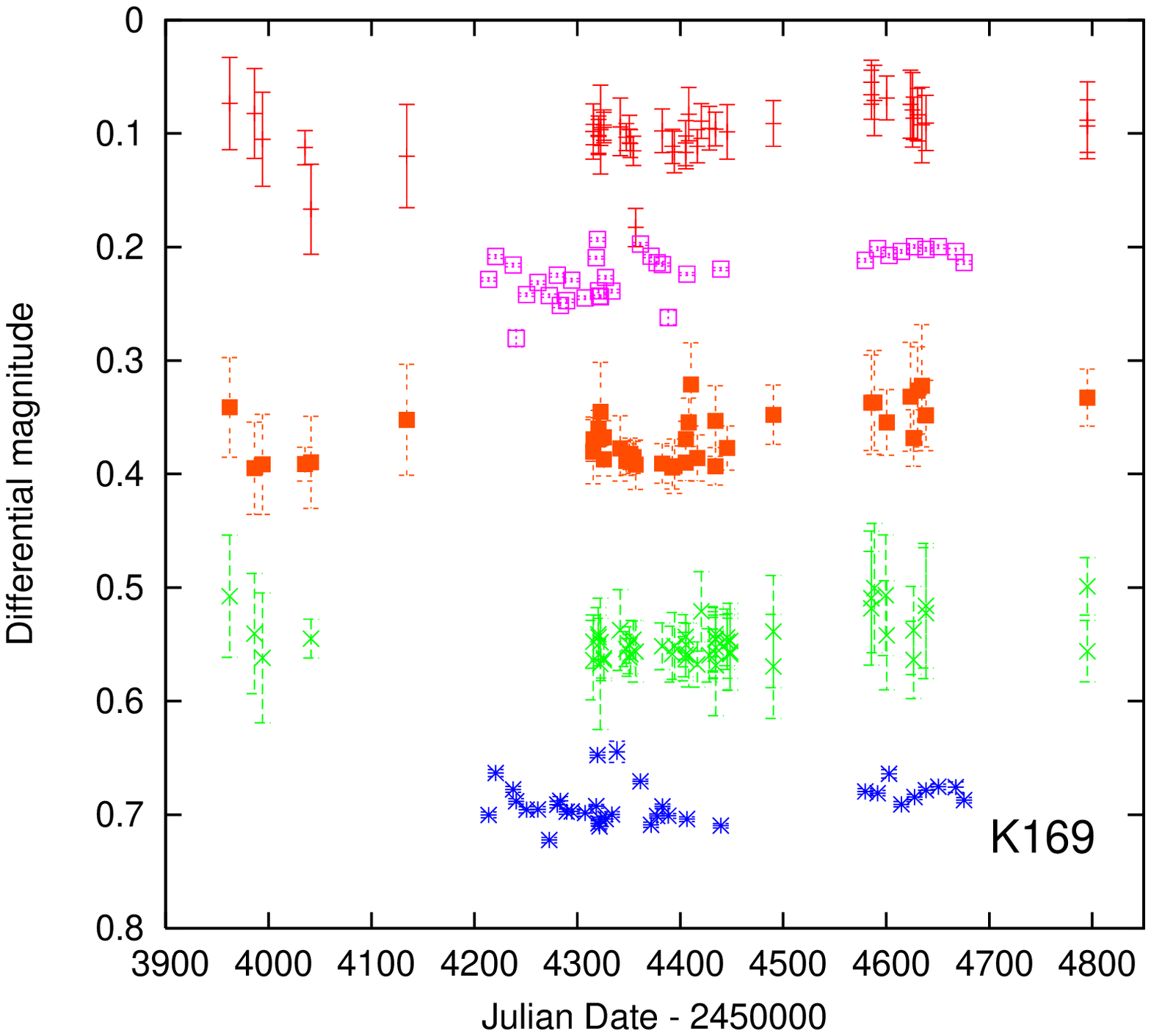}}}
\centerline{\resizebox{\hsize}{!}{\includegraphics[angle=270]{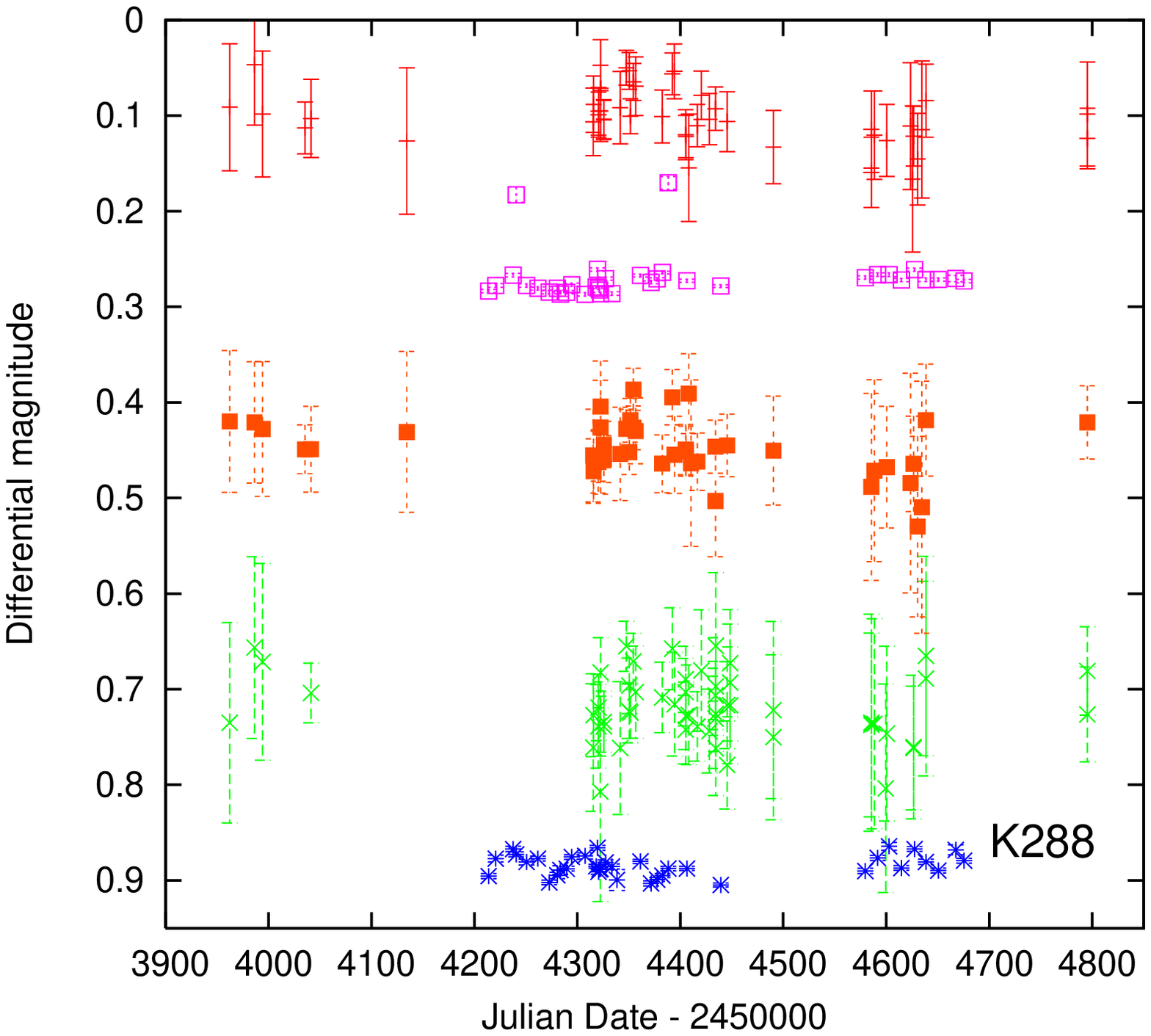}\includegraphics[angle=270]{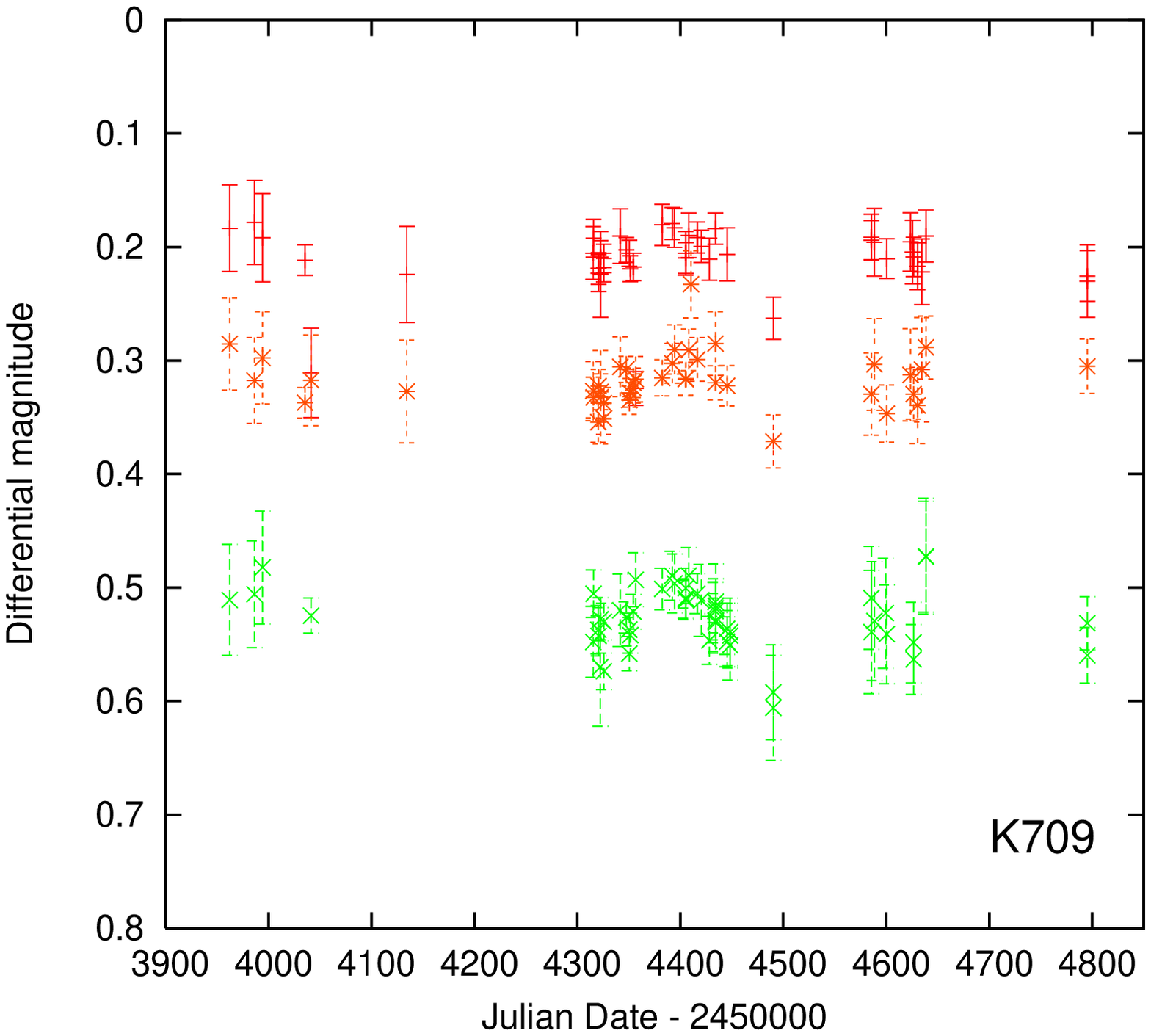}\includegraphics[angle=270]{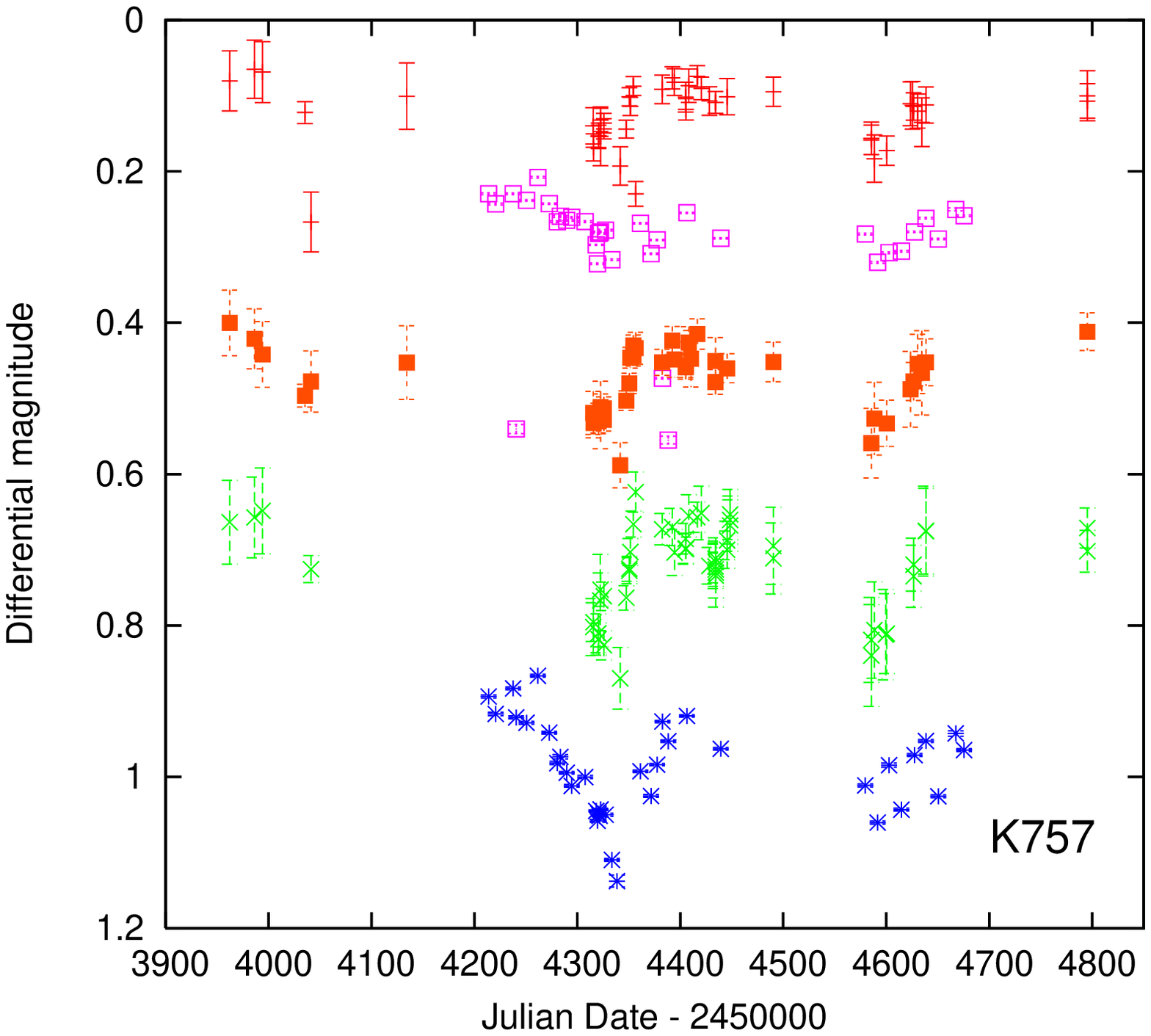}}}
\centerline{\resizebox{0.33\hsize}{!}{\includegraphics[angle=270]{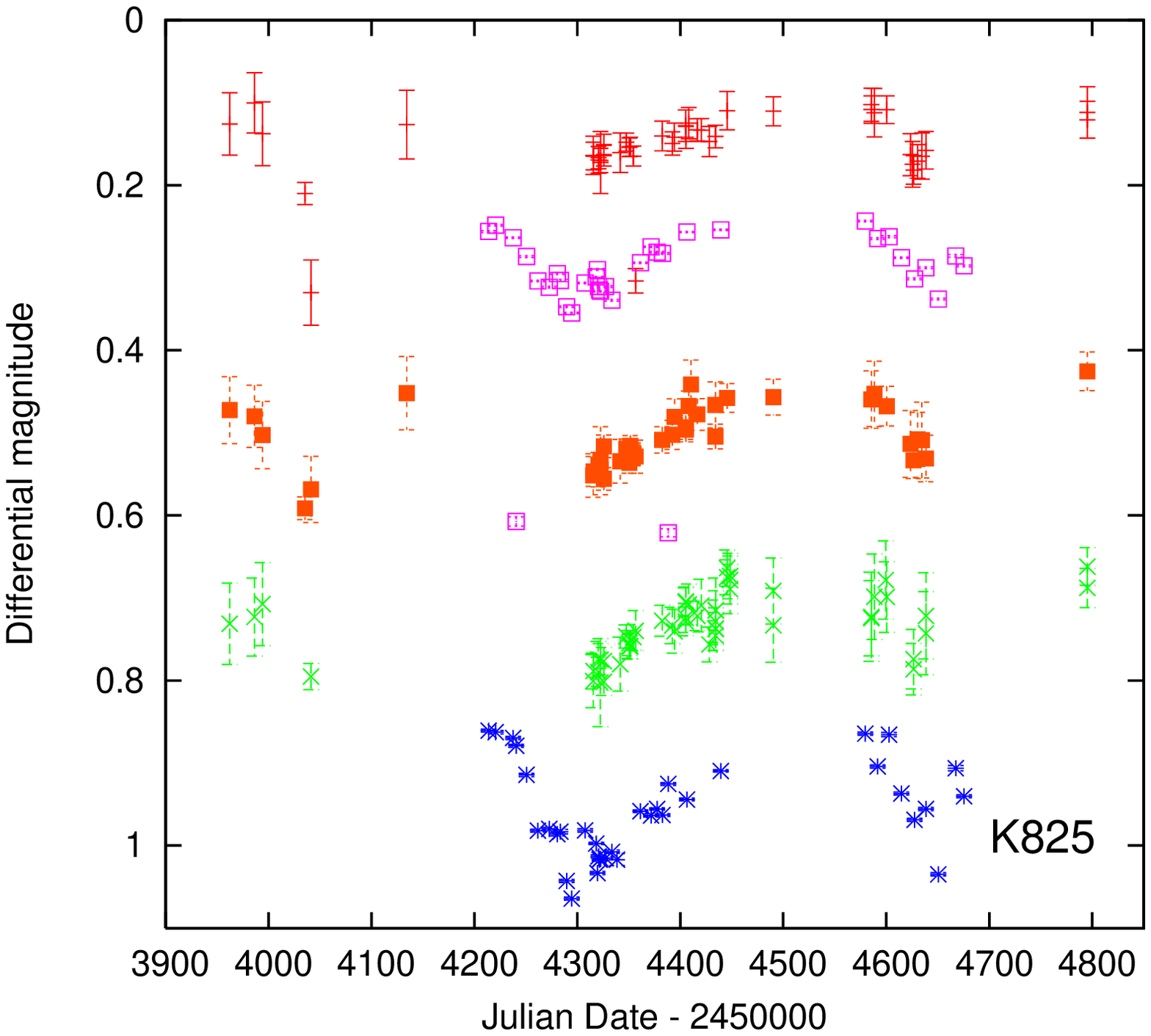}}}
 \caption[Light curves]{Light curves of M15 candidate variables. Data have been offset vertically for clarity. Blue asterisks and magenta hollow squares represent Liverpool Telescope $g^\prime$ and $i^\prime$ photometry, while green crosses, orange filled squares and red dashes show $V$-, $R$- and $I$-band Keele Thornton Telescope data, as labelled in the light curve of K134.}
 \label{M15PhotFig}
\end{figure*}

A similar reduction was performed on the LT data. The field here is much smaller, though the seeing is generally better. A different set of comparison stars (Table \ref{ComparisonTable}) was used, following similar selection criteria. This builds in redundancy against comparison star variation as we have two completely independent data sets.

Fluxes and differential magnitudes were calculated on each processed image. These were summed over each night to improve fidelity. Here again, the non-variability of K238 (Arp II-29) is interesting, as it is suggested to be a comparatively-bright AGB star \citep{FPC83}.

Internal comparisons between comparison stars show possible small-amplitude variations in K497, K508 and K589 on the order of 20--30 mmag. The residual long-term variation in the other six stars is $<$10 mmag. When constructing a summed comparison flux for each image, these three stars were discounted.

The combined results from both the KT and LT are shown in Fig.\ \ref{M15PhotFig}. Differential magnitudes of the target stars are listed in Tables \ref{KTObsTable} \& \ref{LTObsTable}.

\begin{center}
\begin{table*}
\caption[Difference photometry of candidates: Keele]{Difference photometry of comparison stars with variable stars in data from Keele Thornton Telescope. A full electronic table is available online.}
\label{KTObsTable}
\begin{tabular}{l@{}llllllll}

\hline\hline
\multicolumn{2}{l}{Julian}	&	\multicolumn{7}{l}{Apparent photometric magnitude}\\
\multicolumn{2}{l}{Date}	&	K169	&	K825	&	K288	&	K147	&	K757	&	K288	&	K709\\
\hline
\multicolumn{9}{c}{V}\\
\hline
245&3962.43	&	1.408	&	1.231	&	2.385	&	1.145	&	1.464	&	2.554	&	1.211	\\
&	3986.36	&	1.441	&	1.223	&	2.306	&	1.109	&	1.457	&	2.722	&	1.206	\\
&	3994.32	&	1.462	&	1.208	&	2.321	&	1.110	&	1.449	&	2.626	&	1.182	\\
\  & ...	&	...	&	...	&	...	&	...	&	...	&	...	&	...	\\
  \hline
\end{tabular}
\end{table*}
\end{center}


\begin{center}
\begin{table*}
\caption[Difference photometry of candidates: Liverpool]{Difference photometry of comparison stars with variable stars in data from Liverpool Telescope. A full electronic table is available online.}
\label{LTObsTable}
\begin{tabular}{l@{}llllllll}

\hline\hline
\multicolumn{2}{l}{Julian}	&	\multicolumn{6}{l}{Apparent photometric magnitude}\\
\multicolumn{2}{l}{Date}	&	K169	&	K825	&	K288	&	K147	&	K757	&	K288\\
\hline
\multicolumn{8}{c}{$g^\prime$}\\
\hline
245&4213.71	&	2.300	&	1.611	&	2.496	&	1.108	&	1.694	&	2.692	\\
&	4220.69	&	2.263	&	1.613	&	2.477	&	1.092	&	1.717	&	2.666	\\
&	4237.64	&	2.278	&	1.620	&	2.467	&	1.078	&	1.683	&	2.661	\\
\  & ...	&	...	&	...	&	...	&	...	&	...	&	...	\\
  \hline
\end{tabular}
\end{table*}
\end{center}

\subsection{Colour--magnitude diagrams}
\label{M15CMDs}

Colour--magnitude diagrams (CMDs) were generated using {\rm DAOPhot} \citep{Stetson87}. In contrast to the earlier aperture photometry, PSF-fitting was used to extract stellar magnitudes from the images here in order to increase coverage toward the very crowded cluster centre.

The errors calculated from PSF-fitting are correspondingly larger than those from aperture photometry: in the LT data, these are typically $\pm$9--15 mmag for PSF-fitting, compared to $\pm$1--2 mmag for aperture photometry. However, the aperture photometry errors do not take into account close stellar blends: the scatter of magnitudes for non-variable stars suggests that the actual errors in our aperture photometry may be larger, though still smaller than the PSF-fitting errors.

After PSF-fitting, an offset was applied to the pixel co-ordinates from the $i^\prime$ image to match those in the $g^\prime$ image, and the closest object to each detection matched between the bands, within a maximum distance of five pixels (1.3$^{\prime\prime}$).

\begin{figure}
\centerline{\resizebox{\hsize}{!}{\includegraphics[angle=270]{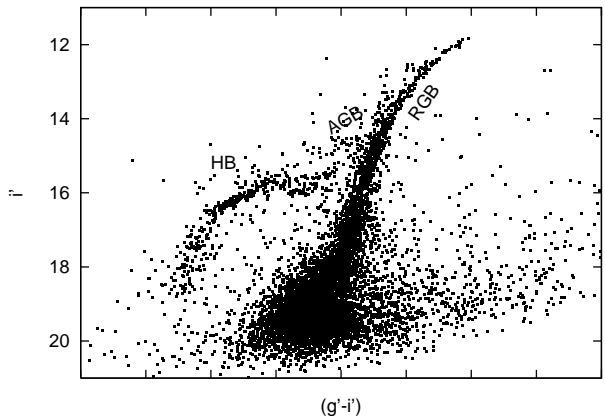}}}
 \caption[Colour-Magnitude Diagram of M15 (LT)]{Colour-magnitude diagram of M15 from the Liverpool Telescope for the night of 2007 October 03. The asymptotic and red giant branches (AGB/RGB) and horizontal branch (HB) are identified, the main sequence turnoff is at $i^\prime \approx 19$ mag.}
 \label{M15LTCMD}
\end{figure}

Due to the lack of literature observations with $g^\prime$-, $i^\prime$- and $R$-band magnitudes, and lack of good colour transformations to obtain these from $B$- $V$- and $I$-band data alone, we have no absolute flux measurements to calibrate the \linebreak $g^\prime$-, $i^\prime$- and $R$-bands. To circumvent this, we have computed theoretical AB magnitudes of K238 in these bands, using the Sloan DSS filter transmissions\footnote{http://www.sdss.org/dr3/instruments/imager/\#filters} and the spectral energy distribution (SED) parameter-estimation technique used in \cite{MvLD+09}, to give an approximate physical magnitude. We find the star has a temperature of 4583 K and luminosity of 805 L$_\odot$, assuming $E(B-V) = 0.10$ mag \citep{Harris96}, and calculate $g^\prime$ = 13.783 mag, $i^\prime$ = 12.506 mag and $R$ = 12.532 mag. Note these values may be in error by $\sim$0.1 mag. The final CMD is shown in Fig.\ \ref{M15LTCMD}.

The CMD clearly resolves the red giant branch as a well-denoted sequence extending from a clearly-defined RGB-tip at $i^\prime \sim 12$ mag, down to the main-sequence turnoff at $i^\prime \approx 19$ mag. The horizontal branch (mainly around $i^\prime \sim 16$ mag) can also be seen extending to very blue colours. The dip at around $(g^\prime - i^\prime) = -0.3$ mag has also been found in earlier studies (e.g.\ \citealt{Stetson94}), as has the kink and gap at the very top of the giant branch ($i^\prime \approx 12.3$ mag). Much of the scatter away from the giant and horizontal branches comes from poor magnitude determination or cross-matching near the crowded cluster centre.

\subsection{Variability}
\label{VarSect}

Although variability is clearly seen in Fig.\ \ref{M15PhotFig} in several stars, the strength, periodicity and regularity of variation is difficult to determine due to the long-period nature of the variations.

In order to determine where variability occurs, we have created the index $\mu$. We calculate this as follows: for each star and filter, we take the stellar magnitude time series and remove any points that deviate from the last observation by more than 0.15 mag. This removes any bad data deviating by $\gtrsim 5 \times$ the standard photometric error from the average observation. For this clipped dataset, we take the standard deviation, $\sigma_{\rm S}$ and divide it by the standard deviation of the point-to-point differences, $\sigma_{\rm P}$, giving $\mu = \sigma_{\rm S} / \sigma_{\rm P}$. For a non-variable star, $\mu$ should be very near-unity. Slight departures from unity may be due to remaining bad data, but values significantly above unity should be due to variability. The index is equally sensitive to both regular and irregular variability.

For each star, we have listed in Table \ref{MuTable} two variability statistics: $\mu_{\rm K}$ for the KT, given by the average of $\mu$ in the $V$-, $R$- and $I-$bands; and $\mu_{\rm L}$ for the LT, which is the average of $\mu$ in the $g^{\prime}$ and $i^{\prime}$ bands. We also give the r.m.s.\ variation, which demonstrates the stability of the KT data: the only star with a higher r.m.s.\ in these data is K288, which is close to the core and suffers from some blending. Finally, we also include the Pearson product-moment correlation co-efficient, $r$, between binned KT and LT data (positive values approaching unity denote good correlation). Here, each telescope's data is represented by a 30-day bin of the deviation from average magnitude, over all filters. This assists differentiation between intrinsic variability and variability caused by changing observing conditions.

\begin{center}
\begin{table}
\caption[Variability indices of candidate variables]{Variability indices of candidate variables, for the KT ($\mu_{\rm K}$) and LT ($\mu_{\rm L}$), respectively. Values of $\mu>1$ suggest variability, the derivation is described in the text. ``r.m.s.'' denotes the average root mean square variation over all filters from the KT ($V$, $R$, $I$) and LT ($g^\prime$, $i^\prime$), respectively. The value $r$ is the correlation co-efficient between the KT and LT data.}
\label{MuTable}
\begin{tabular}{l@{\qquad}c@{\quad}c@{\qquad}c@{\quad}c@{\qquad}c@{\qquad}l}
    \hline \hline
\ 	& \ 		& \ 		& \multicolumn{2}{l}{r.m.s.}	& \ 	& Variability\\
ID	& $\mu_{\rm K}$	& $\mu_{\rm L}$	& \multicolumn{2}{l}{\llap{(}mmag)}& $r$& detected?\\
    \hline
K134	& 1.19		& ---		& 30	& ---	& ---	& No \\
K147	& 1.36		& 2.23		& 15	& 24	&--0.35	& Indefinite$^\ast$ \\
K169	& 1.14		& 0.94		& 18	& 20	&--0.36	& No \\
K288	& 1.28		& 1.33		& 31	& 11	&--0.40	& No \\
K709	& 1.35		& ---		& 19	& ---	& ---	& Maybe \\
K757	& 1.65		& 2.13		& 44	& 46	& +0.73	& Yes \\
K825	& 1.51		& 1.98		& 34	& 41	& +0.91	& Yes \\
\hline
\multicolumn{4}{l}{$^\ast$ See \S\ref{K147Sect} for details.}\\
\hline
\end{tabular}
\end{table}
\end{center}

From Table \ref{MuTable}, we can see that both telescopes show high values of $\mu$ --- indicating the presence of variability --- in K757 and K825. Marginal detections of variability are possible in K709, for which there is no LT data, and K147. In the latter case, however, there appears no correlation between the trends observed from the KT and those from the LT, so we neither claim nor refute variability here.

\section{Discussion}

\begin{figure*}
\centerline{\resizebox{\hsize}{!}{\includegraphics[angle=270]{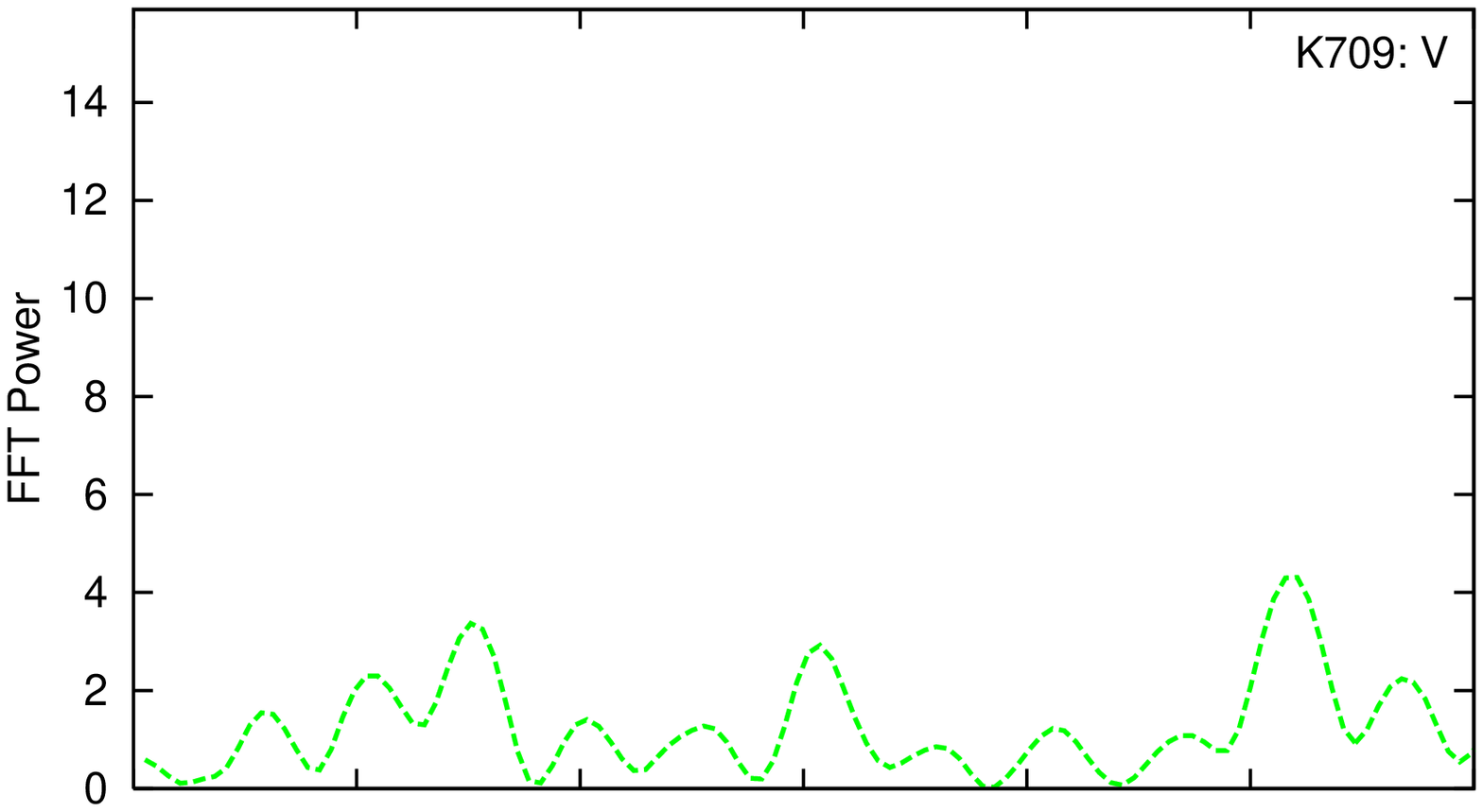}\includegraphics[angle=270]{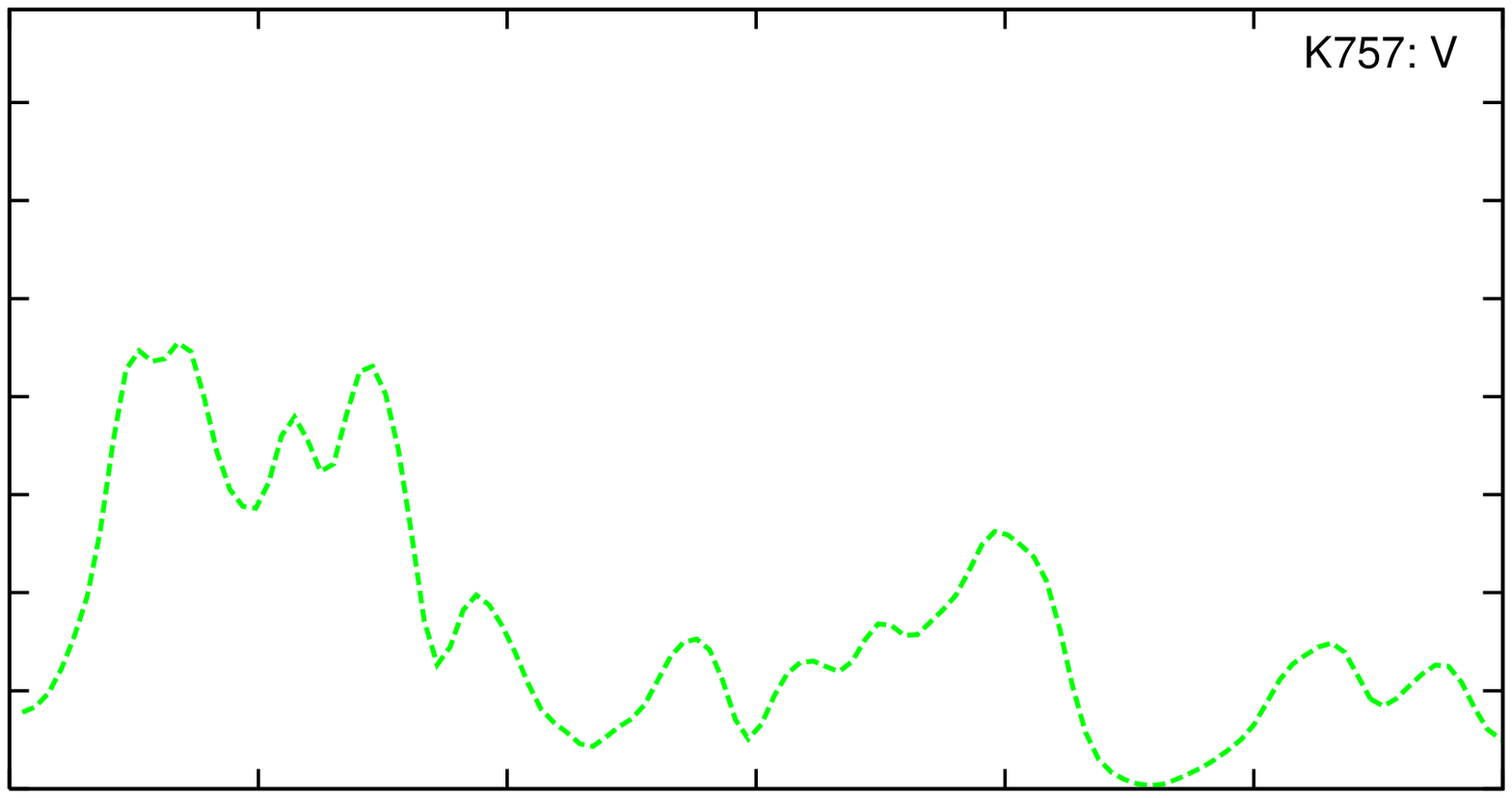}\includegraphics[angle=270]{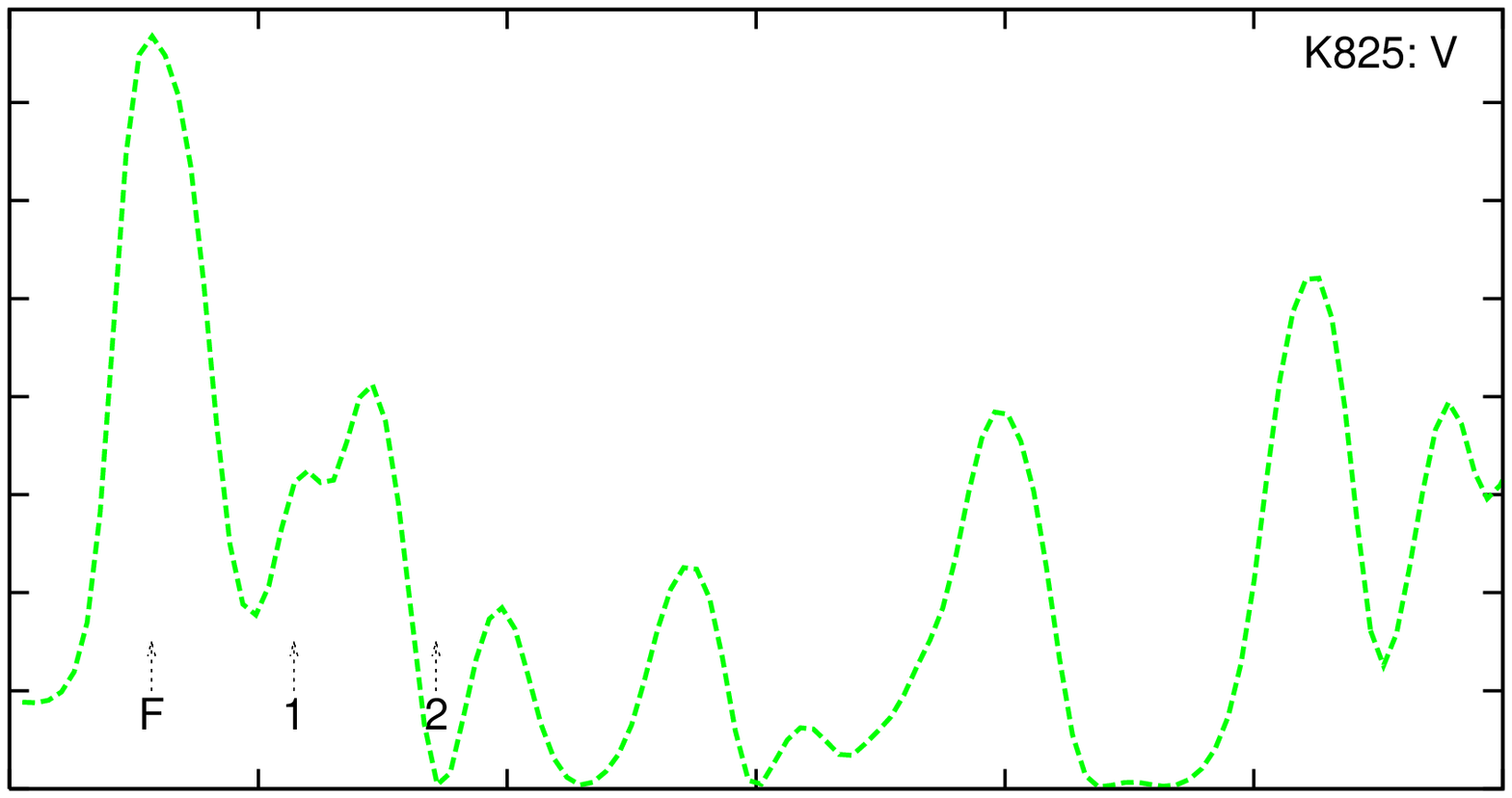}}}
\centerline{\resizebox{\hsize}{!}{\includegraphics[angle=270]{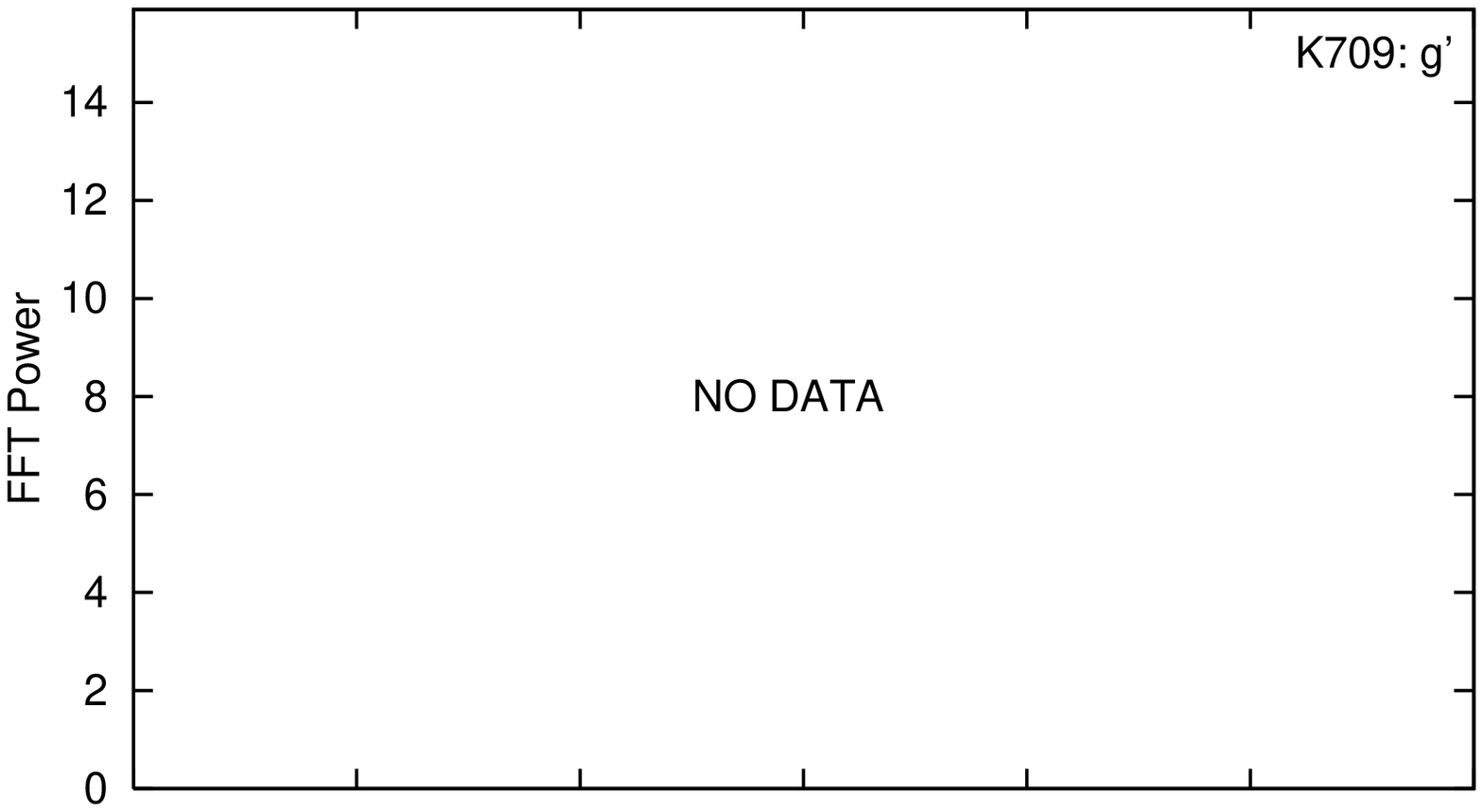}\includegraphics[angle=270]{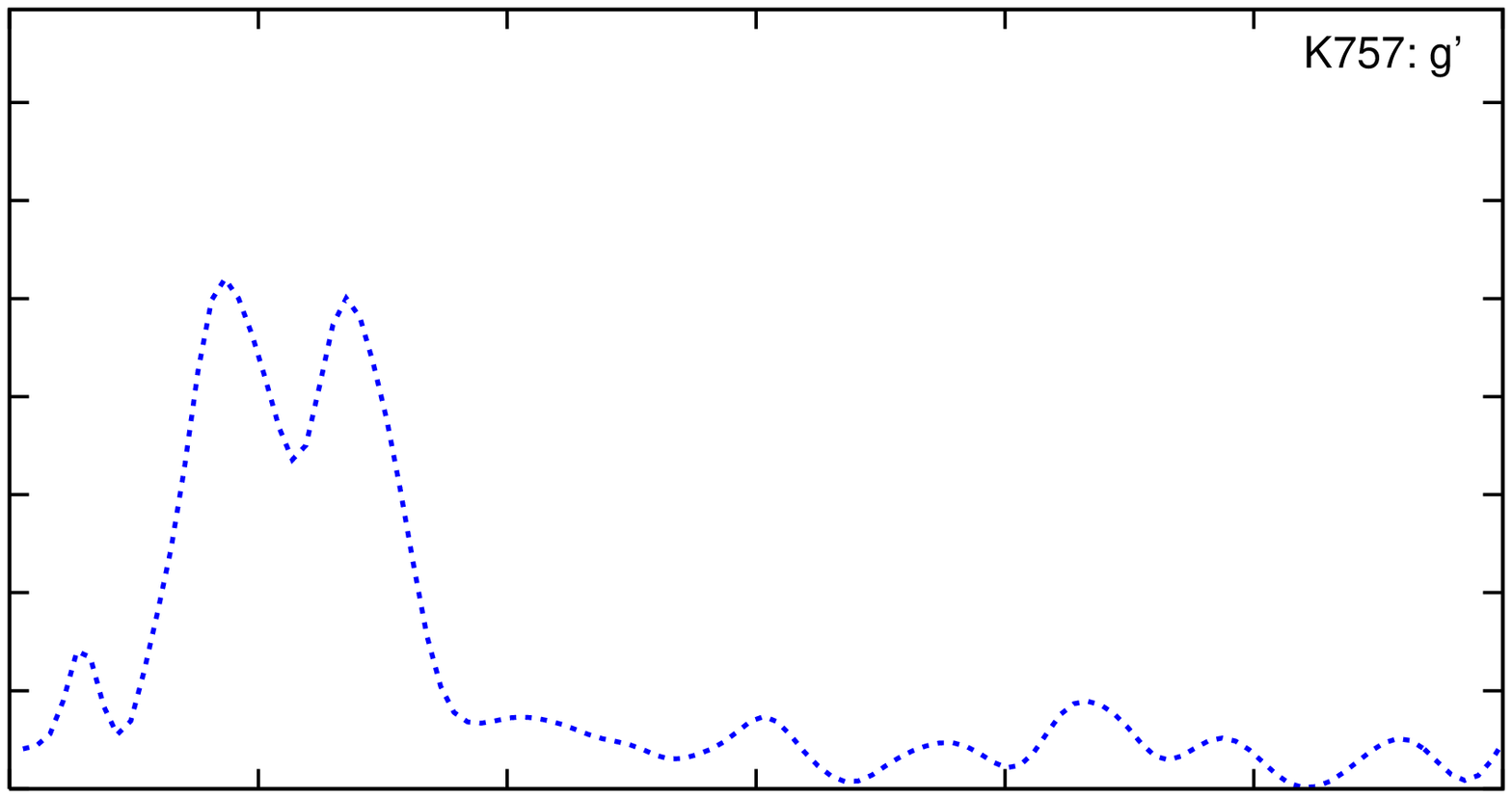}\includegraphics[angle=270]{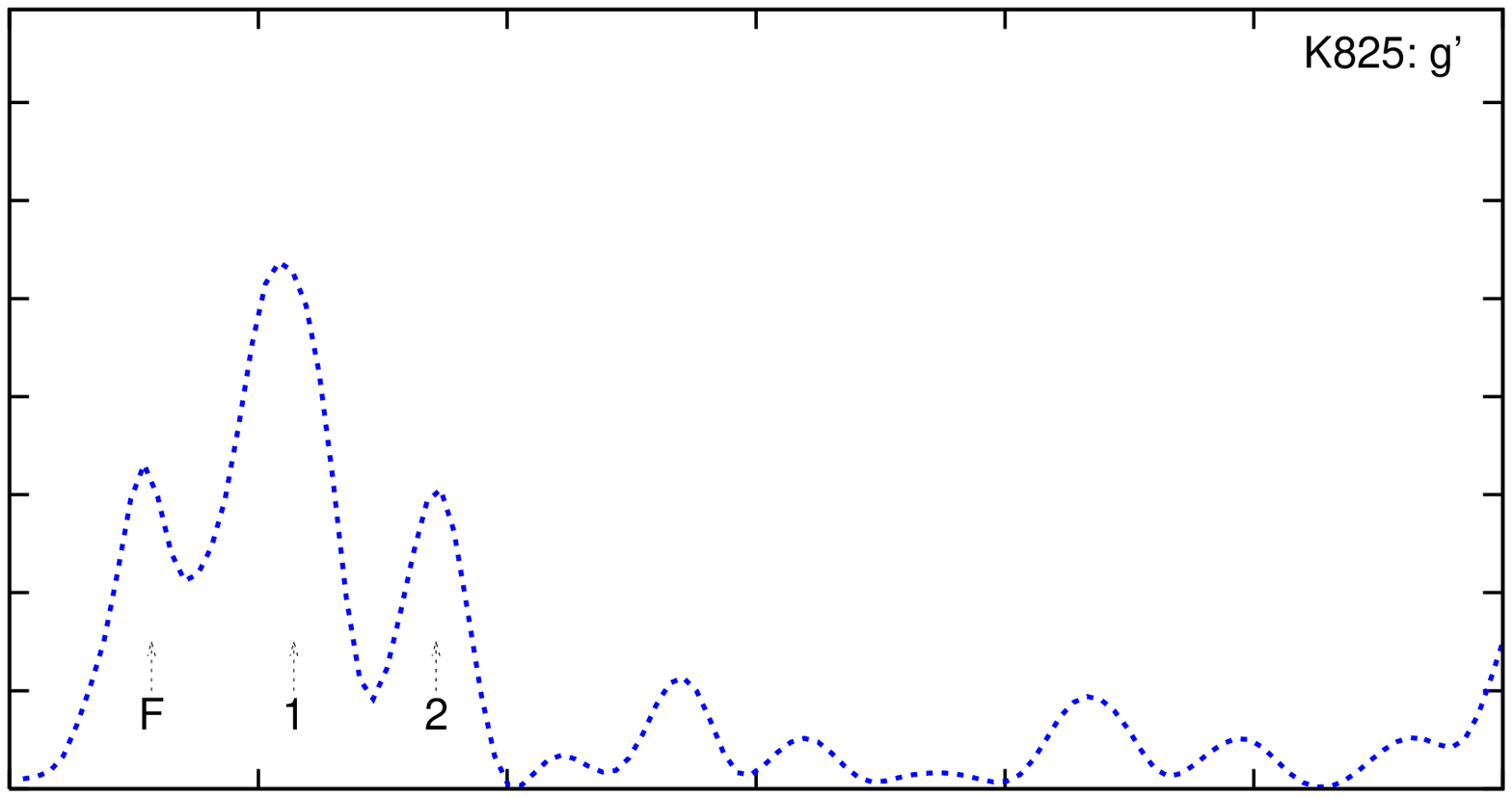}}}
\centerline{\resizebox{\hsize}{!}{\includegraphics[angle=270]{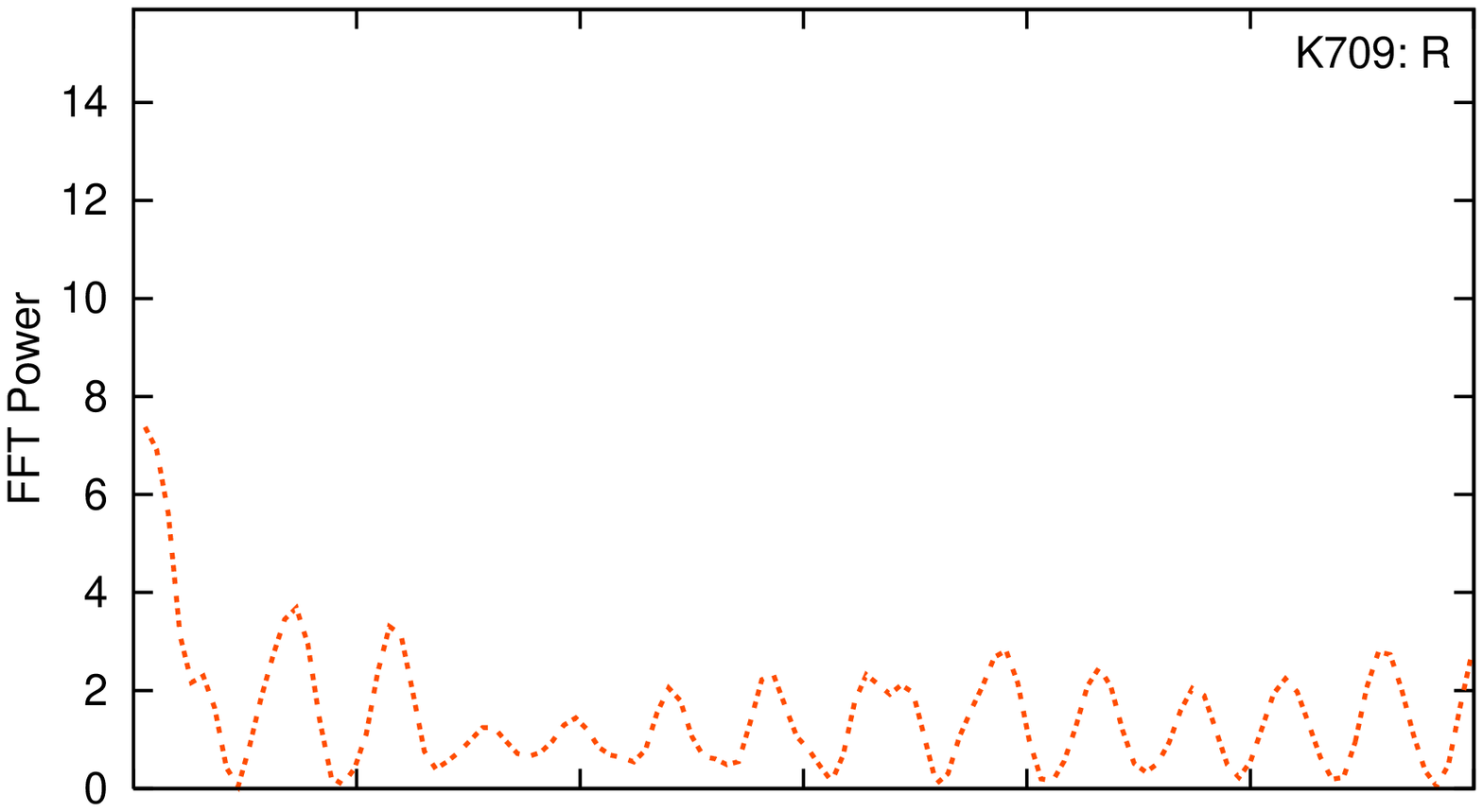}\includegraphics[angle=270]{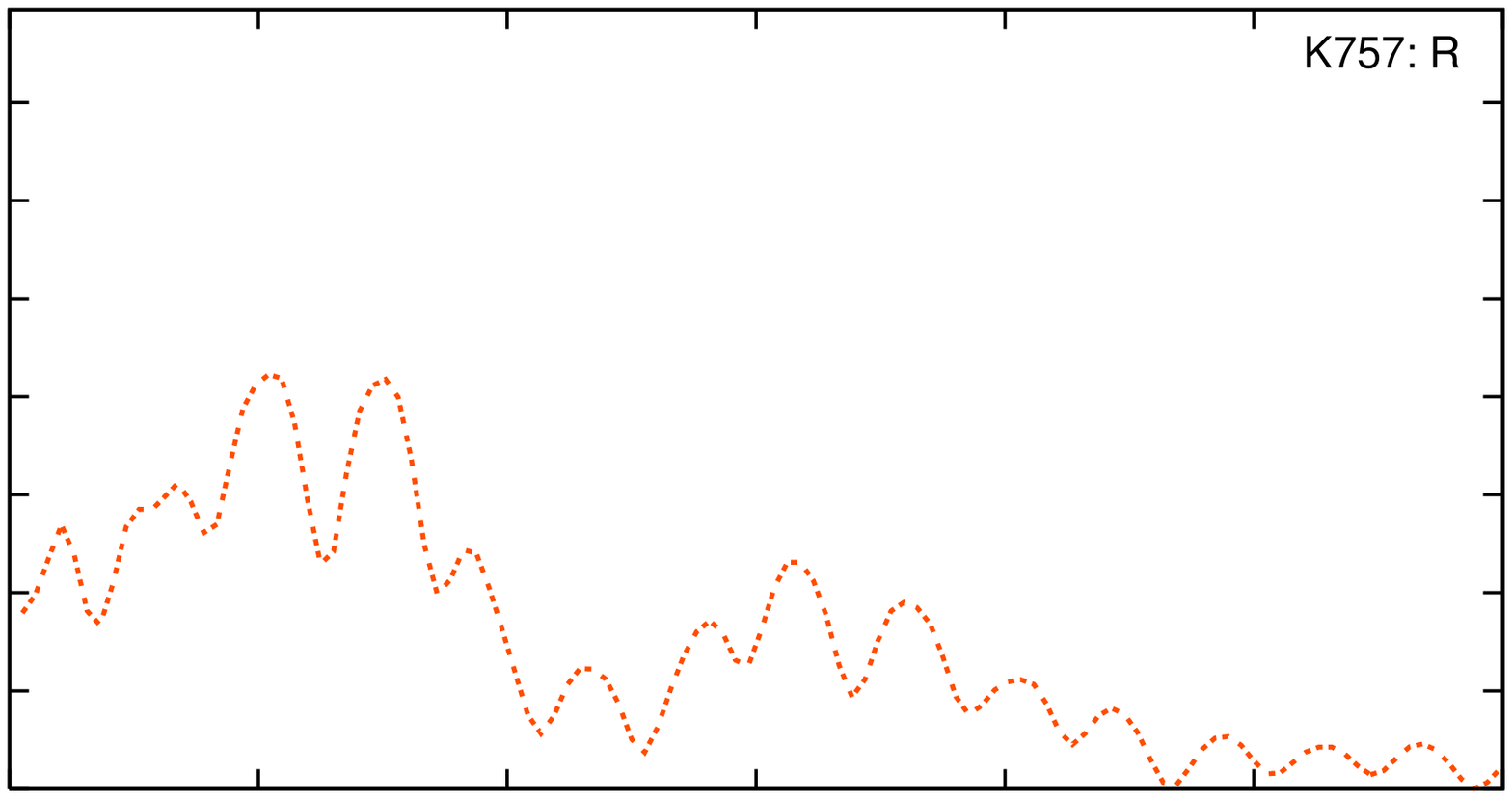}\includegraphics[angle=270]{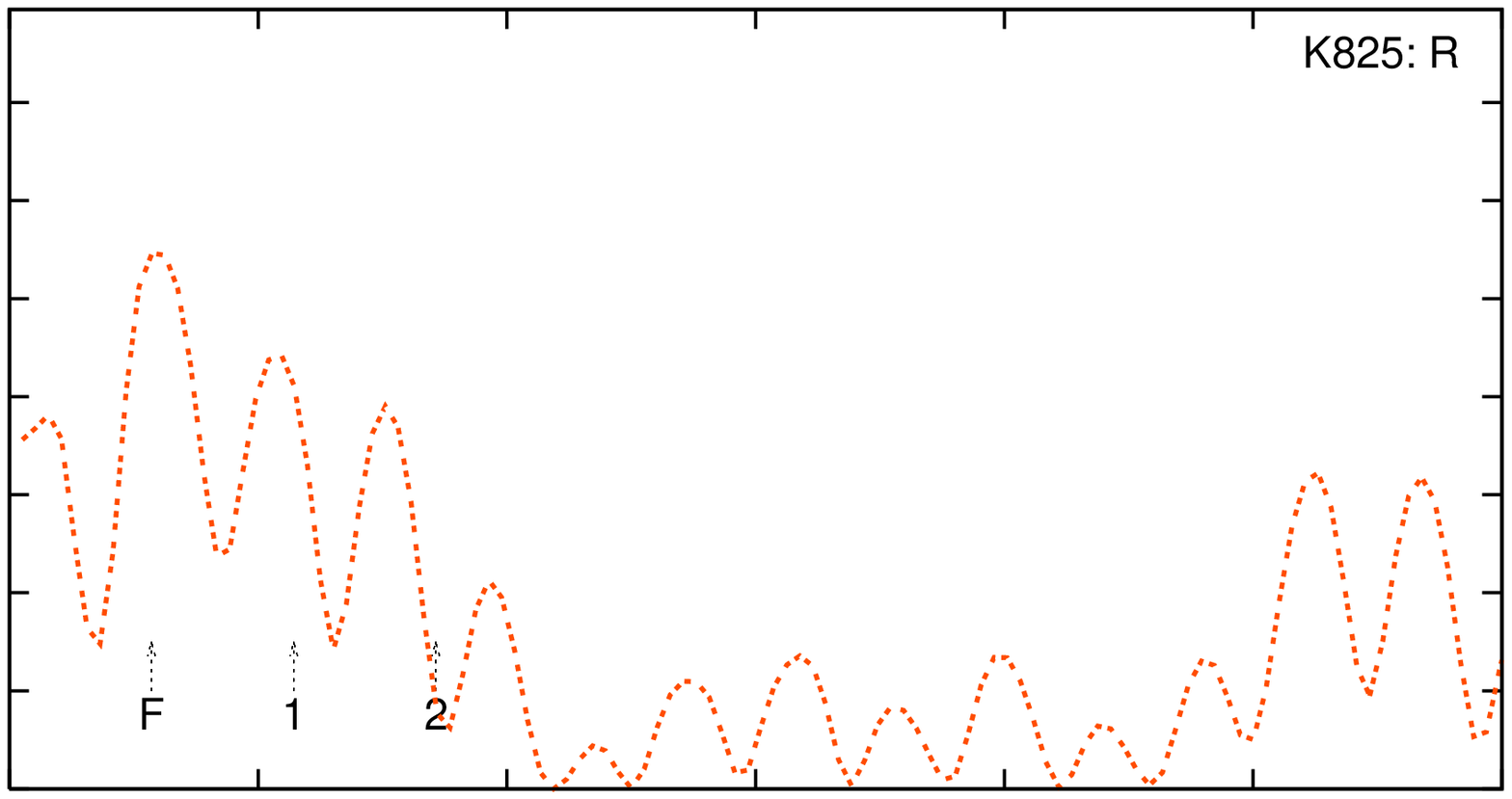}}}
\centerline{\resizebox{\hsize}{!}{\includegraphics[angle=270]{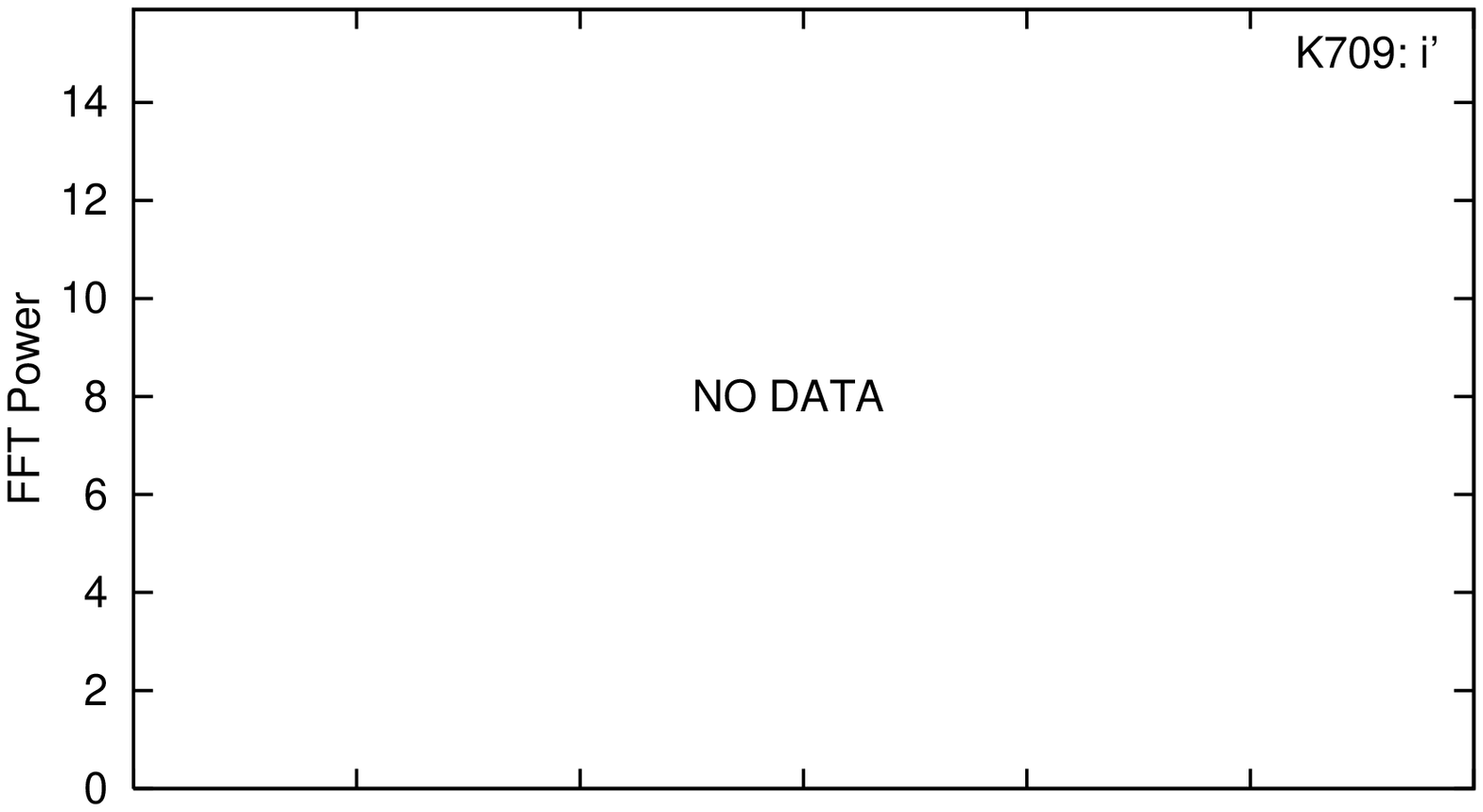}\includegraphics[angle=270]{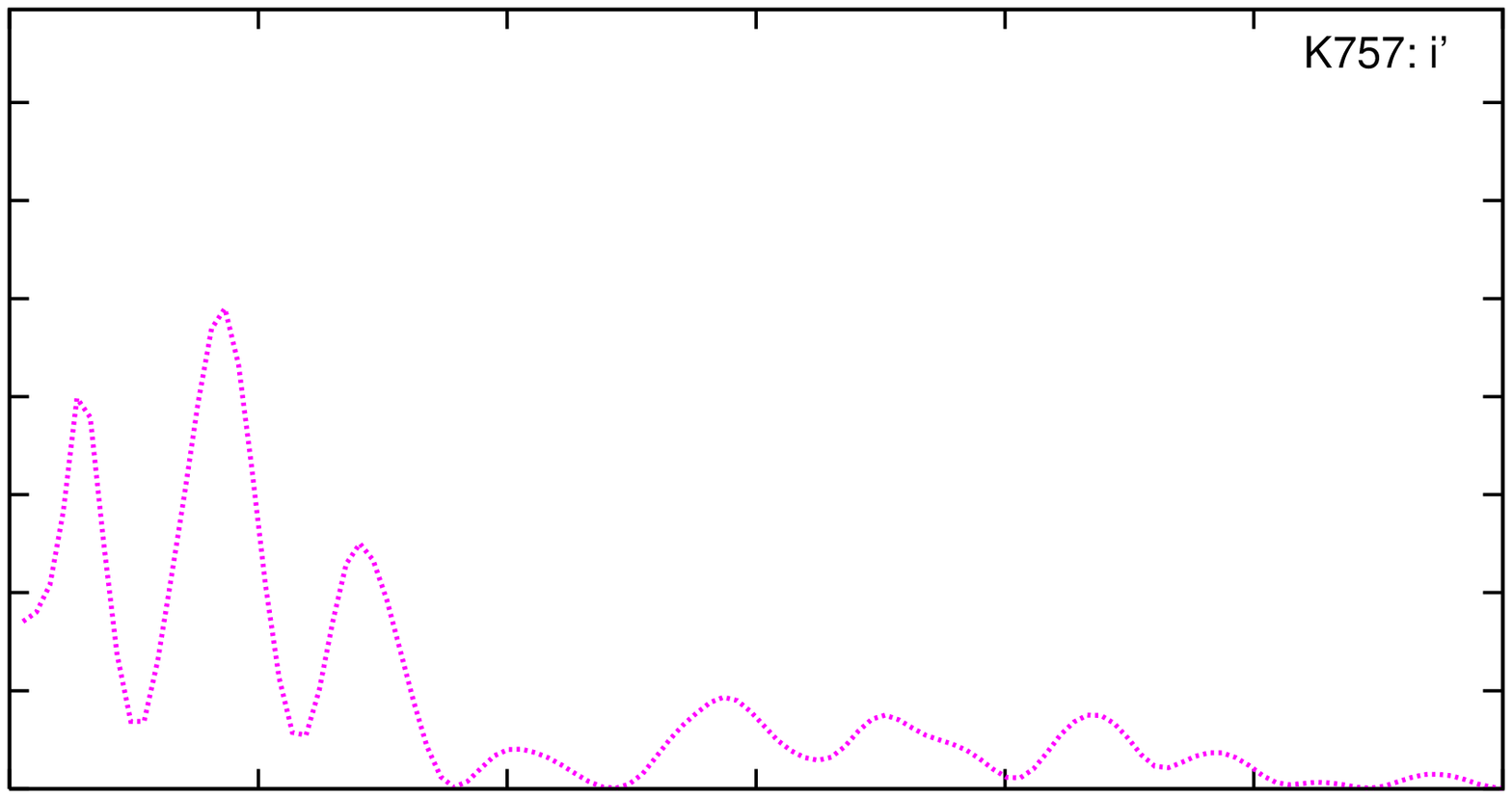}\includegraphics[angle=270]{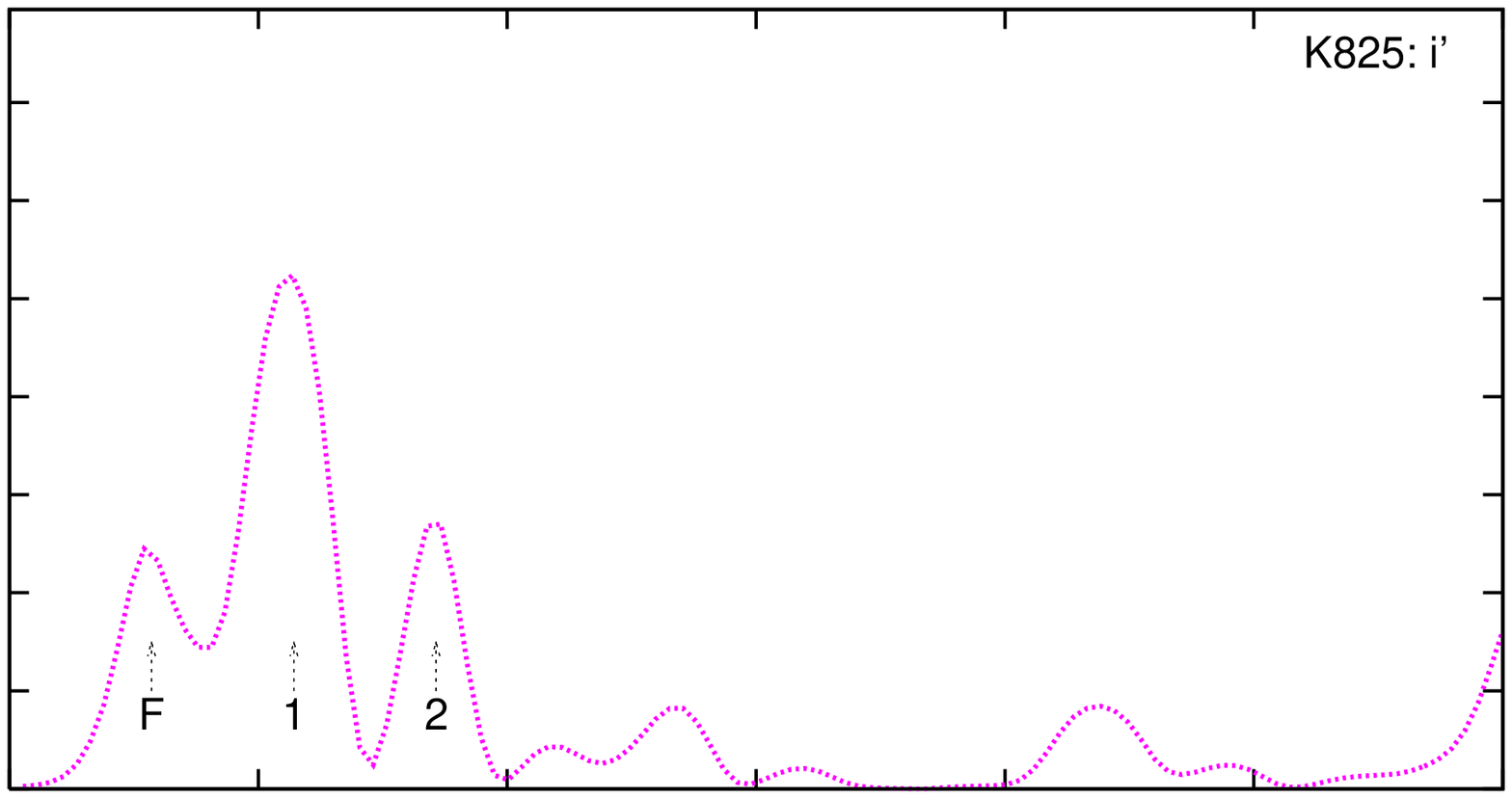}}}
\centerline{\resizebox{\hsize}{!}{\includegraphics[angle=270]{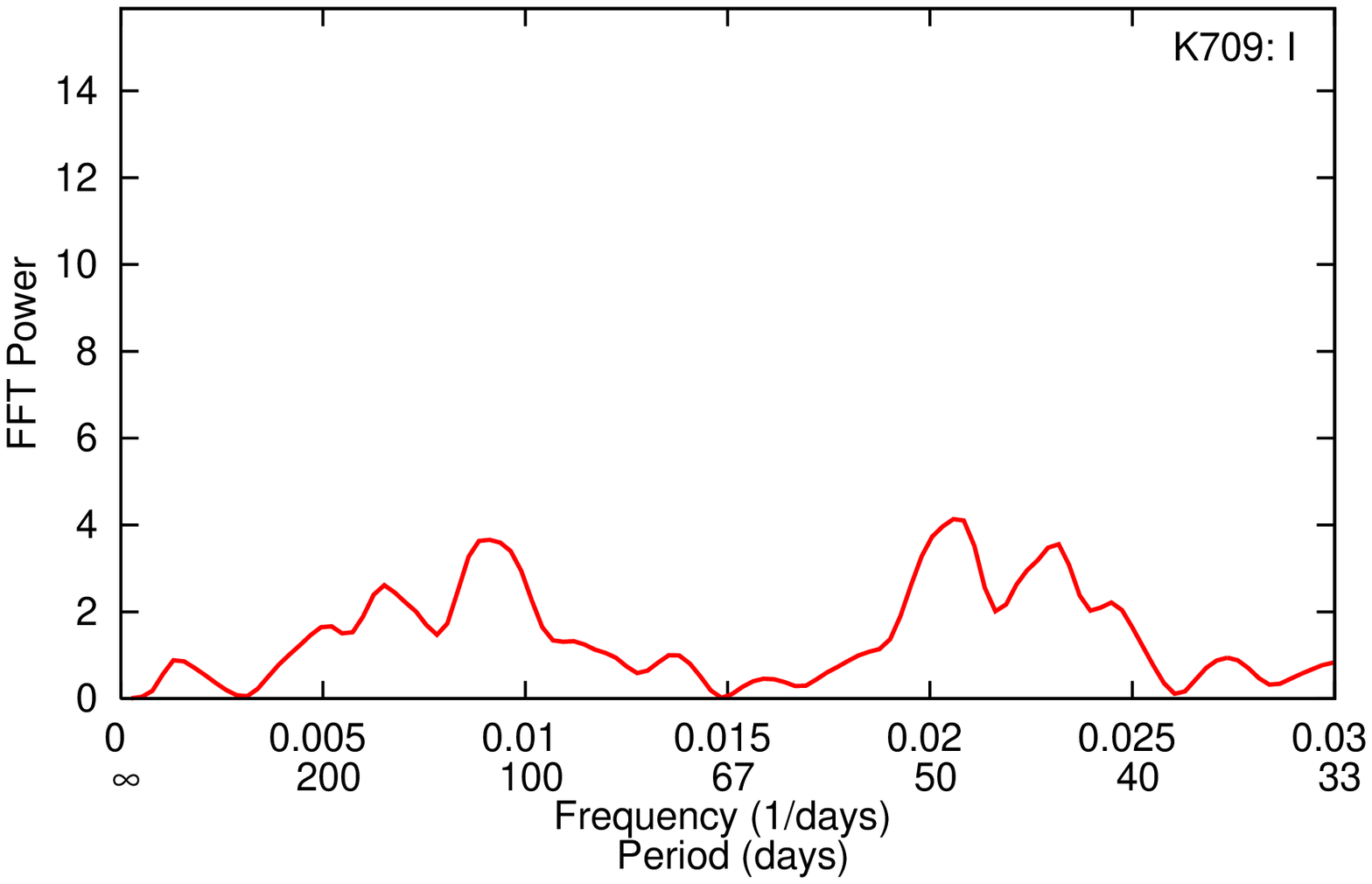}\includegraphics[angle=270]{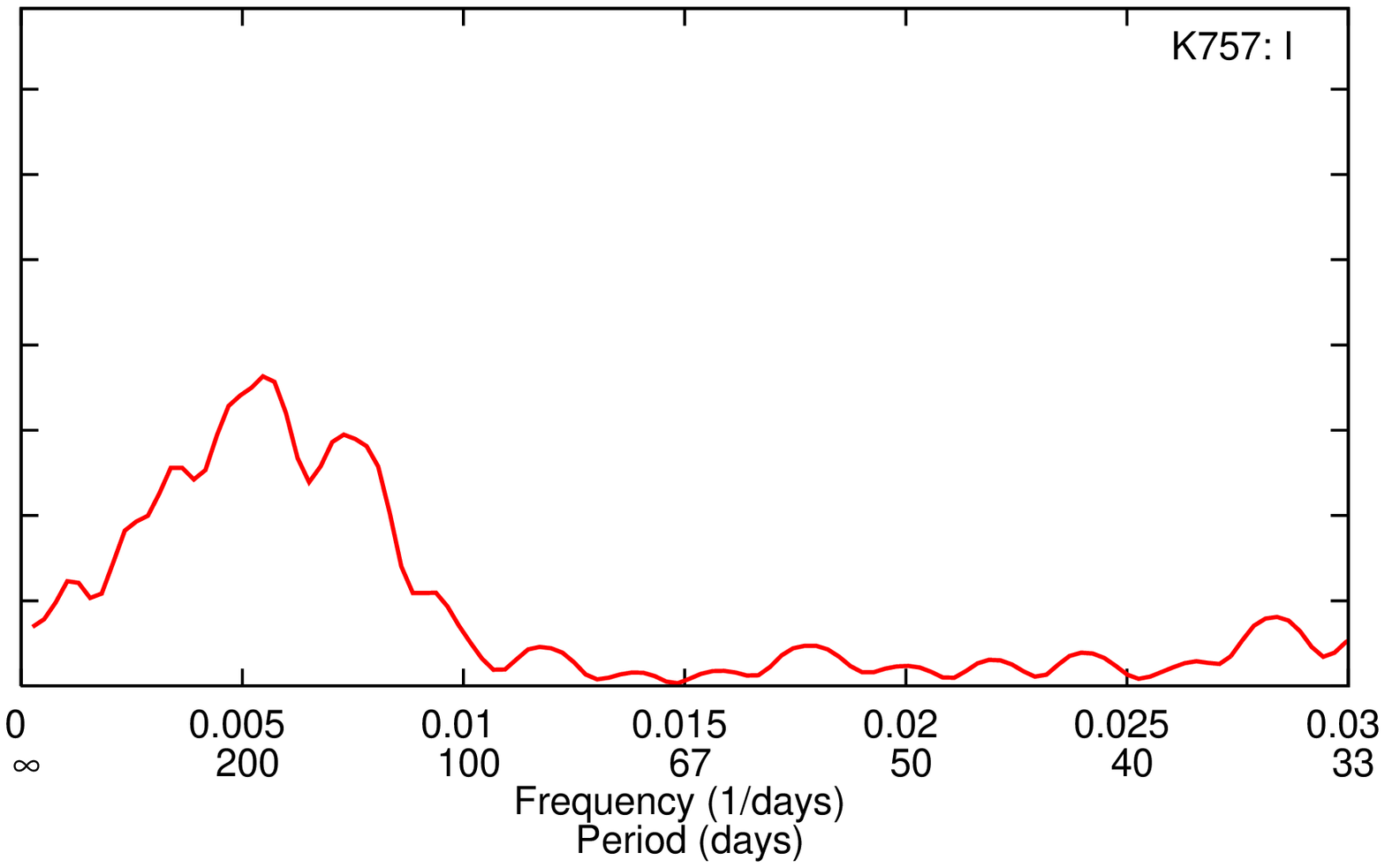}\includegraphics[angle=270]{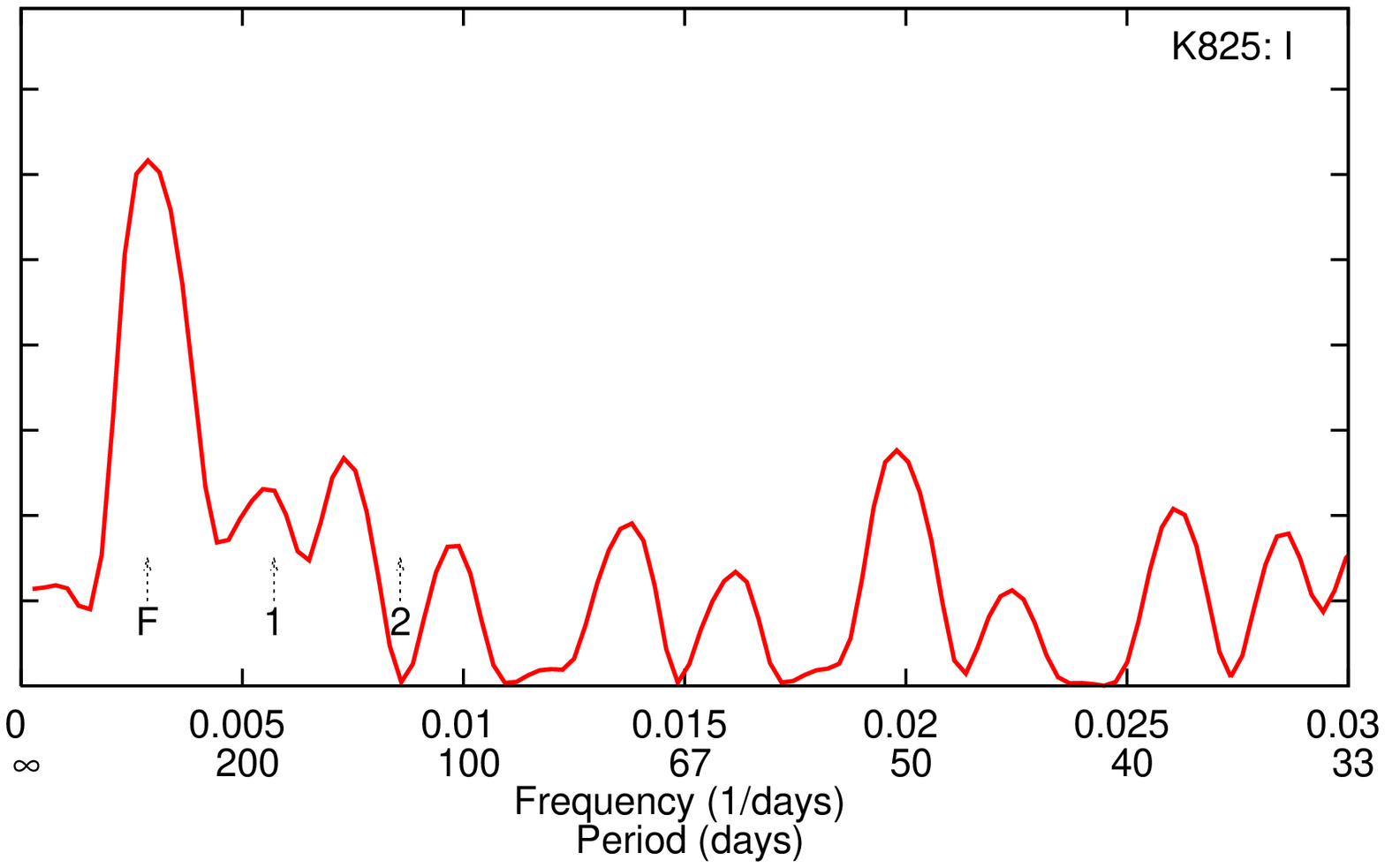}}}
 \caption[FFTs of lightcurves]{Fast fourier transforms of the optical photometry. The colour scheme is as Fig.\ \ref{M15PhotFig}, namely: (top to bottom) V-, g$^\prime$-, R-, i$^\prime$- and I-bands, respectively. Arrows for K825 show the proposed fundamental mode of 350 days (F), and its first (1) and second (2) harmonics.}
 \label{M15FFTFig}
\end{figure*}

\subsection{Individual stars}

The results for individual stars are summarised in Table \ref{M15CudworthTable}.

\subsubsection{K134 (Arp III-8)}

Although not always in the field of view of the LT, there is nevertheless sufficient data from both telescopes to make two independent tests of variability in this star. There appears to be no variation above the scatter in the data, and we can thus conclude that K134 does not show any substantial long period variations. Note that this star is not thought to be a cluster member \citep{Cudworth76}.

\subsubsection{K147 (Arp III-34)}
\label{K147Sect}

One of the brightest stars in the cluster, K147 is a relatively blue star: \citet{BBC+83} show it to have a near-zero ($U-B$) colour. Its proper motion and radial velocity show it to be a field star (Table \ref{M15CudworthTable}). While our observations show some variability may be present (\S\ref{VarSect}), data from the two telescopes are discordant about the trend of this variability, suggesting the observed variability is spurious. It is possible that these variations in our data are due to seasonal instrumental effects. The sensitivity of our data means that we cannot rule out variability in this star at low ($\lesssim \pm$1\%) amplitudes.

\subsubsection{K169 (Arp II-64)}

The photometry of K169 has a relatively low scatter. It clearly shows no large-amplitude variations. Our photometry does not suggest variability in this star and we therefore conclude that it does not pulsate significantly on timescales from days to up to a year.

\subsubsection{K288 (Arp II-16)}

K288 shows relatively low-noise photometry in the LT data, but shows a larger scatter in the KT photometry due to its proximity to the cluster core. \citet{Welty85} could not determine variability in this star because of this, though our data show low enough scatter to determine that there appears to be no coherent variation in this star of $\delta V \gtrsim 0.01$ mag. We conclude that this star is also not variable on long timescales.

\subsubsection{K709 (Arp IV-58)}

K709 may show variability of $\delta V \sim 0.08-0.10$ magnitudes, with maxima at JD 2\,454\,400 and 2\,454\,600, and a minimum near JD 2\,454\,500. The star also has a moderately-high variability index in the KT data, comparable with that of K147 (\S\ref{VarSect}). We do not have data from the LT to determine if this variability is real (it lies beyond the LT field of view), and the variations are close to the noise level in the KT data. Variability in this star would be interesting, as \citet{Welty85} find this star to be non-variable, with an r.m.s. of 0.054 mag: the variations seen in our data would seem to be slightly larger. It is possible that Welty missed variability here due to sparsity of data coverage.

Fig.\ \ref{M15FFTFig} shows a Fast Fourier Transform (FFT) in each filter. The left-most peak near 0 cycles day$^{-1}$, most obvious in K709 in the $R$-band, is the timebase of our photometry, but K709's power spectrum shows no other peaks of statistical significance. The data therefore appear to be of insufficient coverage to determine any periodicity, should it exist.


\subsubsection{K757 (Arp IV-38)}

As one of the brightest and reddest stars in the cluster, and also as a confirmed radial velocity member (Table \ref{M15CudworthTable}), this star is a prime candidate for being an LPV. Indeed, the data from both telescopes shows a concurrent dimming around JD 2\,454\,340 for $\sim$100 days. This drop is wavelength-dependent, showing a clear increase with decreasing wavelength, decreasing from $\sim$0.2 magnitudes for the $g^\prime$ data to $<$0.05 magnitudes in $I$-band. Further, smaller drops may be seen near JD 2\,454\,100 and JD 2\,454\,600, though they are less-well covered.

The light curve is very similar to a typical binary star showing a single eclipse and ellipsoidal variations, though the star would have to be in a contact or near-contact binary to show variation at this (approximate) period. We consider it more likely that this star is showing pulsation. 

Again, the data are of insufficient coverage to determine whether a regular period exists to this variability. The FFT (Fig.\ \ref{M15FFTFig}) shows a broad peak in all bands (except $i^\prime$) between 100 and 500 days, though visual examination of the lightcurve (Fig.\ \ref{M15PhotFig}) suggests that it is probably $\sim$250 days. Further data are needed to see if this variation is truly periodic.

\subsubsection{K825 (Arp I-12)}
\label{K825Sect}

K825 is another particularly bright and red star in the cluster which is also a confirmed radial velocity member (Table \ref{M15CudworthTable}). Physically, the star is also the most-luminous star in the cluster, and one of the coolest. While we cannot distinguish between AGB and RGB stars at this luminosity, K825 is likely to define the RGB tip. It is also a known H$\alpha$ variable (see \S\ref{Halpha} and \citealt{MDS+08}), suggesting either chromospheric activity or pulsation.

Variations from this star were immediately visible in the data from an early stage of the analysis. A clear dip is seen near JD 2\,454\,300 and another near JD 2\,454\,650 in all five bands. A further possible dip at JD 2\,454\,050 may be present, but there is only one epoch of observation here. Again, these dips are wavelength-dependent, decreasing from $\sim$0.17 mag in $g^\prime$ to $\sim$0.08 mag in $I$-band.

The light curve shows a classic long-period variation attributable to stellar pulsation, with a rapid drop in brightness at bluer wavelengths as the star expands and cools, then a slower recovery as the star contracts and warms. FFT analysis (Fig.\ \ref{M15FFTFig}) shows peaks in all bands between 330 and 370 days. The strong peak in the LT data at 200 days is an alias of this caused by insufficient data coverage. The lack of a clear drop at JD 2\,453\,950 suggests that the amplitude of variability is irregular and thus that this star is an SRV.

\subsection{Variability in the context of stellar evolution}

\subsubsection{Pulsation as the source of variability}
\label{HRDSect}

\begin{figure}
\centerline{\resizebox{\hsize}{!}{\includegraphics[angle=270]{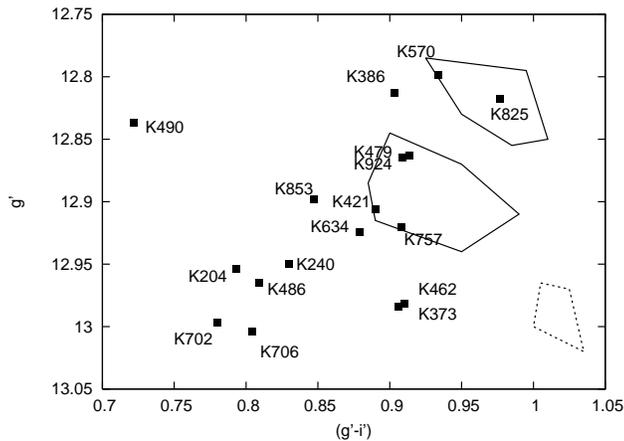}}}
 \caption[Variability in the upper giant branch of M15]{A CMD using the same data as Fig.\ \ref{M15LTCMD}, showing the uppermost part of the giant branch of M15 on 2007 October 03. Individual stars are labelled following the nomenclature of \citet{Kustner21}. The thick lines denote the observed regions within which K757 and K825 varied during the Liverpool Telescope observing runs, the dashed line shows the variation of the non-variable K157 for comparison.}
 \label{M15LTCMD3}
\end{figure}

\begin{figure}
\centerline{\resizebox{\hsize}{!}{\includegraphics[angle=270]{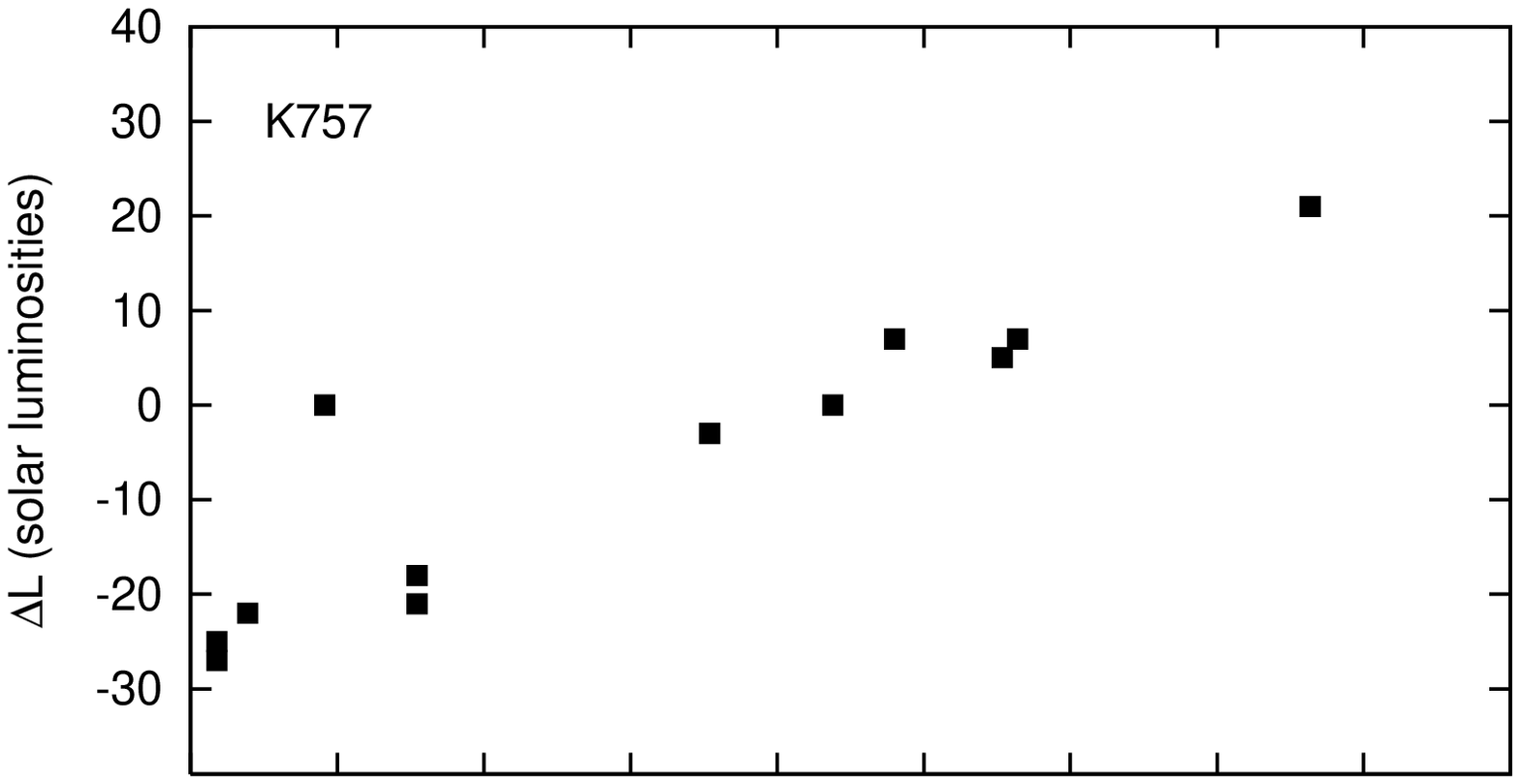}}}
\centerline{\resizebox{\hsize}{!}{\includegraphics[angle=270]{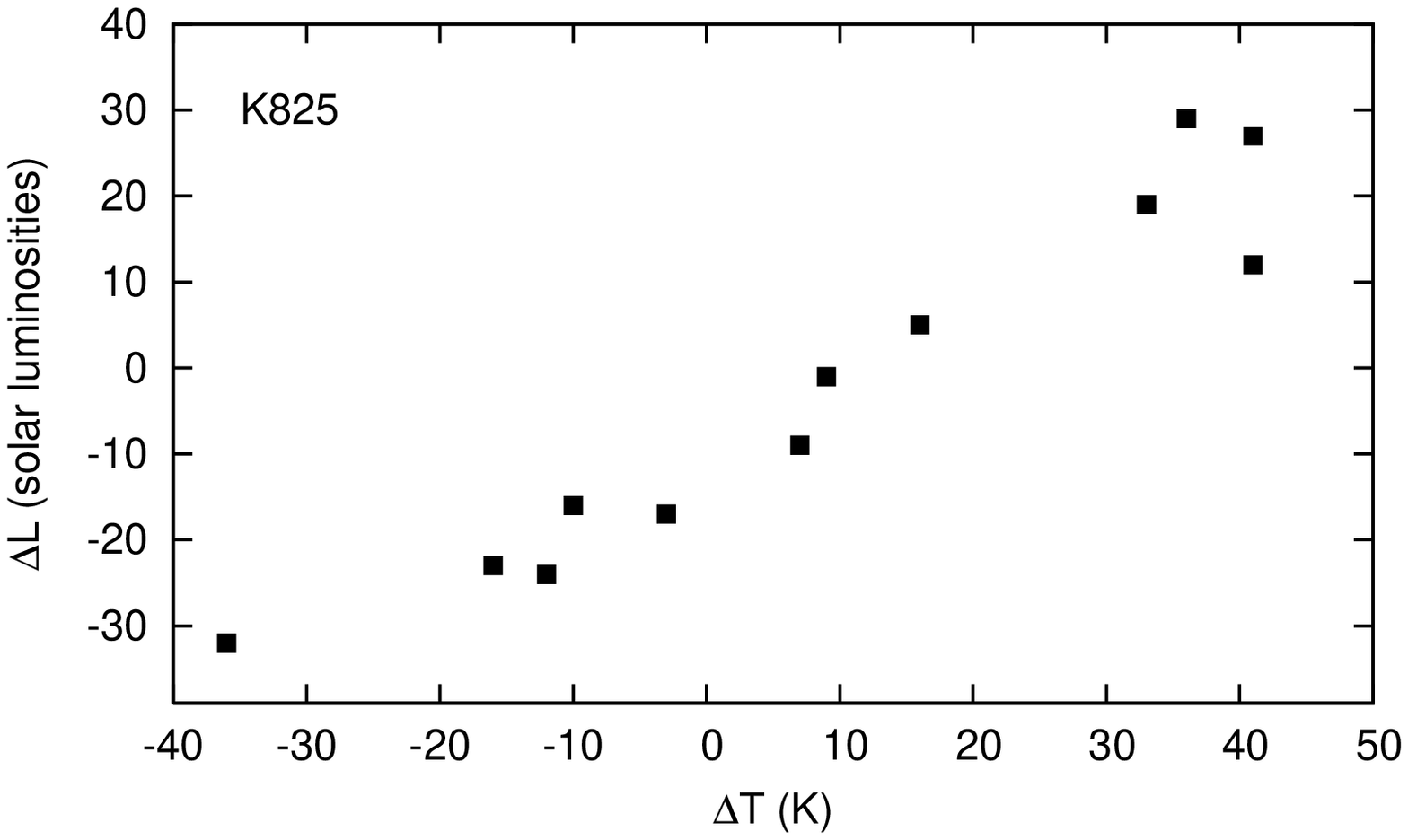}}}
\vspace{2mm}
 \caption[Variability in physical units]{Variation of K757 (top panel) and K825 (bottom panel) in physical units, as determined from SED fitting.}
 \label{M15HRDVarFig}
\end{figure}


We have found variability in two of the cluster's most evolved members (K757 and K825). These stars seem to share the metallicity of the cluster: both are within 0.12 dex of the the cluster's average value, [Fe/H] = --2.4 (\citealt{SKS+97}, 2000). These therefore represent the most metal-poor long-period variables known in our Galaxy. Outside our Galaxy, there is only an unconfirmed member of the Bo\"otes I dwarf galaxy that may be more metal-poor \citep{DOCK+06}. This star has a period of $P \approx 85$ days, $\delta V \approx 0.4$ and an optical magnitude which would place it on the AGB, were it confirmed a member.

Notably, \citet{SPK00} also find both K757 and K825 to be barium-enhanced by [Ba/Fe] $\approx$ +0.37, among the highest enhancement in the cluster, suggesting convective mixing of $s$-process elements (see also \citet{vLvLS+07}). In Sneden et al.'s earlier work, they further suggest that both may be double stars, on the basis of ``weak, blue-displaced extra absorption components''. Two-epoch observations of K757 by \citet{MDS08}, which show a radial velocity shift of 6.2 km s$^{-1}$, suggest this may be the case. If pulsation is present, the ``extra absorption components'' visible in the spectra may be pulsation-induced line doubling, as seen in, e.g., 47 Tucanae \citep{LWH+05}.

The location of K757 and K825 on the giant branch, along with the variation of these stars in observed colours are shown in Fig.\ \ref{M15LTCMD3}, and physical units in Fig.\ \ref{M15HRDVarFig}. It is clear from these figures that not only are these two stars among the most-evolved in the cluster, but that their variability arises from significant changes in both temperature \emph{and} luminosity. 

Stellar parameters for K757 and K825 were calculated using the same method as for K238 (\S\ref{M15CMDs}), with data from various optical catalogues, 2MASS and \citet{BWvL+06}. For K757, this analysis gives a temperature of $4489 \pm 201 K$, log($g$) = 0.746, and $L$ = 1426 $\pm$ 255 L$_\odot$ for K757; for K825, we find $T = 4411 \pm 155$ K, log($g$) = 0.662, $L$ = 1615 $\pm$ 227 L$_\odot$, which are broadly consistent with literature values (\citealt{FPC83}; \citealt{SKS+97}; \citealt{ASA+00}; \citealt{MDS+08}).

Variation in these parameters was estimated by refitting the SED using 50-day averages of the optical ($g^\prime, V, R, i^\prime, I$) magnitudes. Fig.\ \ref{M15HRDVarFig} shows both stars exhibit a variation between a high-temperature, high-luminosity state, and a low-temperature, low-luminosity state, with $\delta T_{\rm eff} \approx \pm50$ K, and $\delta L \approx \pm 30$ L$_\odot$\footnote{Though the absolute errors are large, relative errors can be considered much smaller. Ranges also include a small (estimated $\sim $10 -- 20\%) noise contribution.}. (We presume that, since variability declines with increasing wavelength, there is negligable variation at longer wavelengths than $I$-band.) If we assume that the total optical depth of any dust envelope is largely constant through these changes, and that the stellar spectrum is close to a blackbody, this corresponds to a radius change of $\pm$1.1\% around an average of 62 R$_\odot$ for K757, and $\pm$0.9\% around an average of 69 R$_\odot$ for K825.

This type of variation --- where maximum brightness occurs near-simultaneously for wavelengths longer than $\approx$1 $\mu$m --- is also seen in stronger metal-poor pulsators in globular clusters (cf.\ $\omega$ Cen V42 --- \citealt{DFLE72}; \citealt{MW85}). Similar changes can occur on spotted stars, though a rotation period of 350 days requires a rotation rate of 11 km s$^{-1}$. This is somewhat large for a star this evolved \citep{CSRB+09}. Furthermore, it would require the spots to be relatively constant in size and strength over a period of $\sim$1 year; it also cannot explain the changes seen in the H$\alpha$ line.

A 350-day sinusoidal variation in radius with $\sim$1\% amplitude yields a peak pulsation amplitude of only $\pm$0.1 km s$^{-1}$ at the photosphere for both K757 and K825. This value is very small compared to both the $\sim$60 km s$^{-1}$ escape velocity, and the speed with which we might expect a dusty wind to be accelerated via photon pressure alone. However, the speed may increase as the pulsation travels outward through the more rarified atmosphere.

For comparison, we can calculate the speed one might expect of a dusty wind. Assuming canonical values of 10 km s$^{-1}$ for a 10\,000 L$_\odot$ solar metallicity star, and $v \propto 10^{\rm 0.5 [Fe/H]} L^{0.25}$ (\citealt{vanLoon00}; \citealt{MvLM+04}), we might expect this speed to be $v \approx 450$ m\,s$^{-1}$ if gas-dust coupling is maintained. If gas-dust coupling is not maintained, and a driving mechanism is coupled to \emph{either} the gas or dust, then we might expect the driven medium to escape, and the other to be left behind and fall back onto the star.

\subsubsection{K825: H$\alpha$ line and gaseous mass loss}
\label{Halpha}

\begin{figure*}
\centerline{\resizebox{\hsize}{!}{\includegraphics[angle=0]{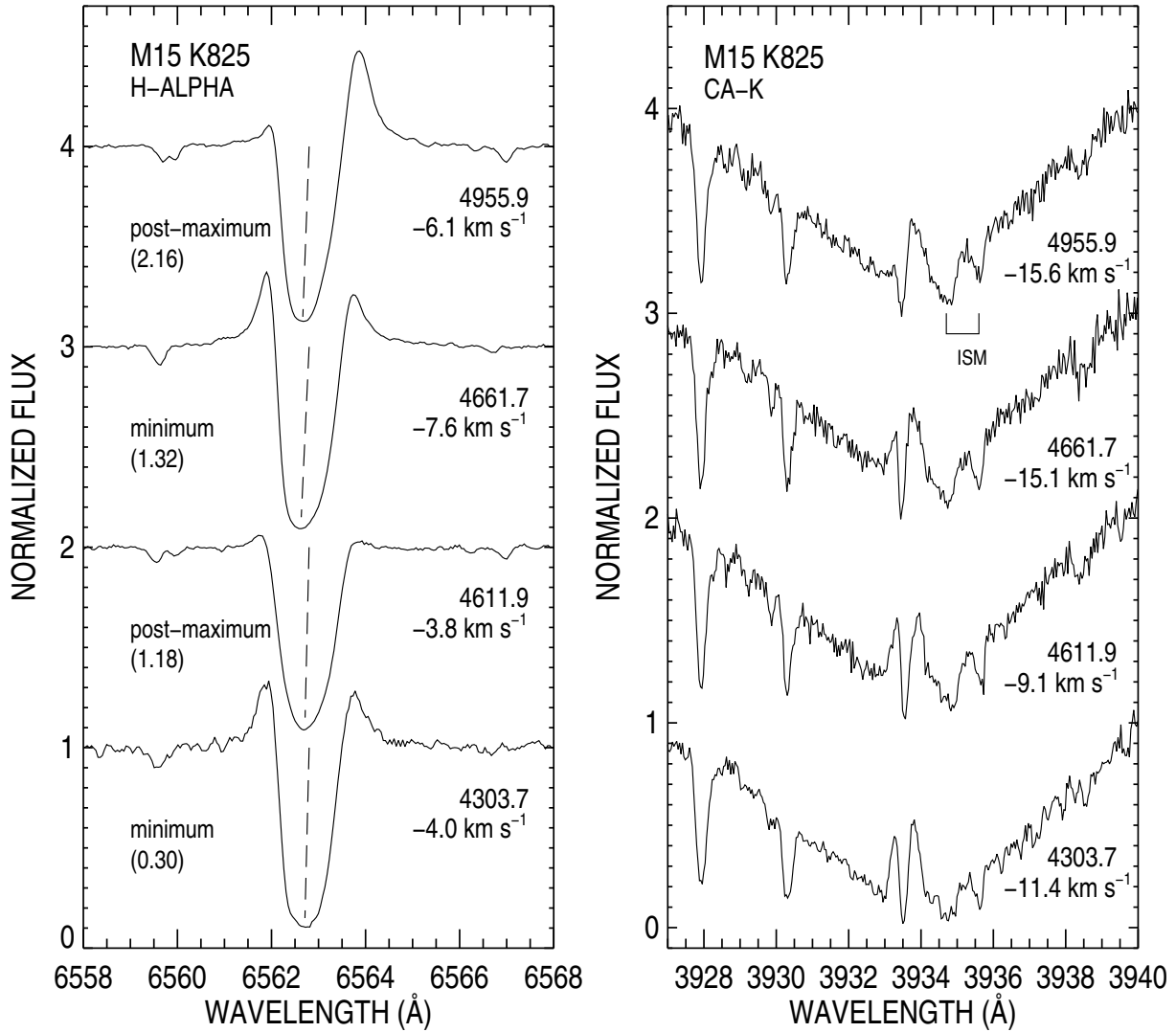}}}
 \caption[H$\alpha$ and Ca II line profiles of K825]{Variation in the H$\alpha$ and Ca \protect{\sc ii} K profiles of K825, as observed with Magellan/MIKE. H$\alpha$ line bisectors are shown as dotted lines. Truncated Julian Dates and the velocity shift in the line cores are shown. Wavelengths are shown at the stellar rest velocity and interstellar medium (ISM) components are identified. Bracketted numbers denote pulsation cycles since TJD 4200.0 (visual maximum), assuming a 350-day period; phases should be considered approximate. Cf.\ \cite{MAD09} for line profile variations in K757.}
 \label{HalphaFig}
\end{figure*}

\emph{Spitzer} photometry \citep{BWvL+06} shows no infrared excess around K757 and K825, suggesting a lack of circumstellar dust production. It appears that these stars are of too low a metallicity to sustain adequate dust production to drive a stellar wind. In the absence of a dust-driven wind, we turn our attention to the stellar chromosphere: blue-shifted absorption cores, often coupled with line emission wings, have been observed in several notable lines and are taken as tracers of the mass-loss rate and velocity of an outflow. Such lines include Mg {\sc ii} k, Na {\sc i} D, Ca {\sc ii} K, H$\alpha$, He {\sc i} and various UV lines (e.g.\ \citealt{DHA84}; \citealt{DWP99}; \citealt{CBR+04}; \citealt{DLY+05}; \citealt{DLS07}; \citealt{MvL07}; \citealt{DSS09}).

We can compare our light curve to recent spectroscopic observations of K825 taken as part of an unrelated programme (K757 has already been presented in \citealt{MAD09}). Fig.\ \ref{HalphaFig} presents spectra taken with the MIKE double-echelle spectrograph mounted on the Magellan/CLAY 6.5\,m telescope at Las Campanas Observatory using a 0.75$^{\prime\prime}$ $\times$ 5$^{\prime\prime}$ slit, giving a resolution of $R \approx 40\,000$. The spectra were reduced with bias subtraction, flat-fielding and sky subtraction using the MIKE-IDL pipeline updated by J.~Prochaska\footnote{http://web.mit.edu/$\sim$burles/www/MIKE/}. Th-Ar arc exposures bracketting the stellar targets were used to determine the wavelength scale. Four epochs are available: two fortuitously near subsequent photometric minima, the other two epochs being shortly after adjecent photometric maxima. For each spectrum, the line bisector is also shown and the velocity shift of the line core with respect to the photospheric rest velocity indicated. These velocities are generally related to the wind outflow velocity but do not necessarily approximate them --- modelling suggests that actual outflow velocities are higher than indicated by the H$\alpha$ profiles (\citealt{MCP06}; \citealt{MvL07}; \citealt{MAD09}). 

Variation seen in our lightcurve of K825 is to some extent repeated in its H$\alpha$ and Ca {\sc ii} K line profiles: the spectra taken at photometric minimum show a moderate asymmetry in the H$\alpha$ emission, while the spectra taken near maximum light show either zero or reversed asymmetry in the H$\alpha$ wings. The Ca {\sc ii} K lines show similar variations in the strength of their emission, though the red emission component is always dominant here. The behaviour is not entirely reproducable between epochs --- there is considerably greater asymmetry at phases 1.32 and 2.16, presumably due to a more massive outflow. This suggests a more chaotic variation of the chromosphere, partially but not entirely coupled to the pulsation period.

It is interesting to note that the H$\alpha$ and Ca {\sc ii} line cores are permanently blue-shifted. There is no obvious correlation with pulsation period here either, though the velocities of H$\alpha$ and Ca {\sc ii} correlate well with each other. The permanency of this blueshift, despite substantial changes in the emission from the (lower altitude) chromosphere, strongly suggests that there is a permanent bulk outflow from the star. As Ca {\sc ii} K absorption occurs higher in the atmosphere than H$\alpha$, and the Ca {\sc ii} blueshifts are larger, we can also state that the wind is strongly accelerating in this region. This means that star's mass loss may be enhanced by pulsation, and there may well be other driving mechanisms (e.g.\ magnetic reconnection emitting Alfv\'en waves) which operate within the line-forming region of the extended atmosphere.

\citet{MAD09} find that K757 shows similarly permanent blueshifts in its H$\alpha$ line cores, despite changes in the line emission from the chromosphere. By modelling the H$\alpha$ line, they imply that K757's gaseous mass-loss rate changed by a factor of $\sim$6 over their observations. Given both stars show a permanent outflow, with velocities that change by a factor of $<2$, we may speculate that the wind driving mechanism remains relatively constant. The strong changes in emission strengths, however, suggest that the rate at which material is injected into the wind is strongly variable. This further suggests that pulsations may vary the mass-loss rate from these systems, but that the kinetic energy that subsequently accelerates the wind may be provided by another, more-constant source.

\subsubsection{K757, K825 and the $P-L$ diagram}

\begin{figure}
\centerline{\resizebox{\hsize}{!}{\includegraphics[angle=0]{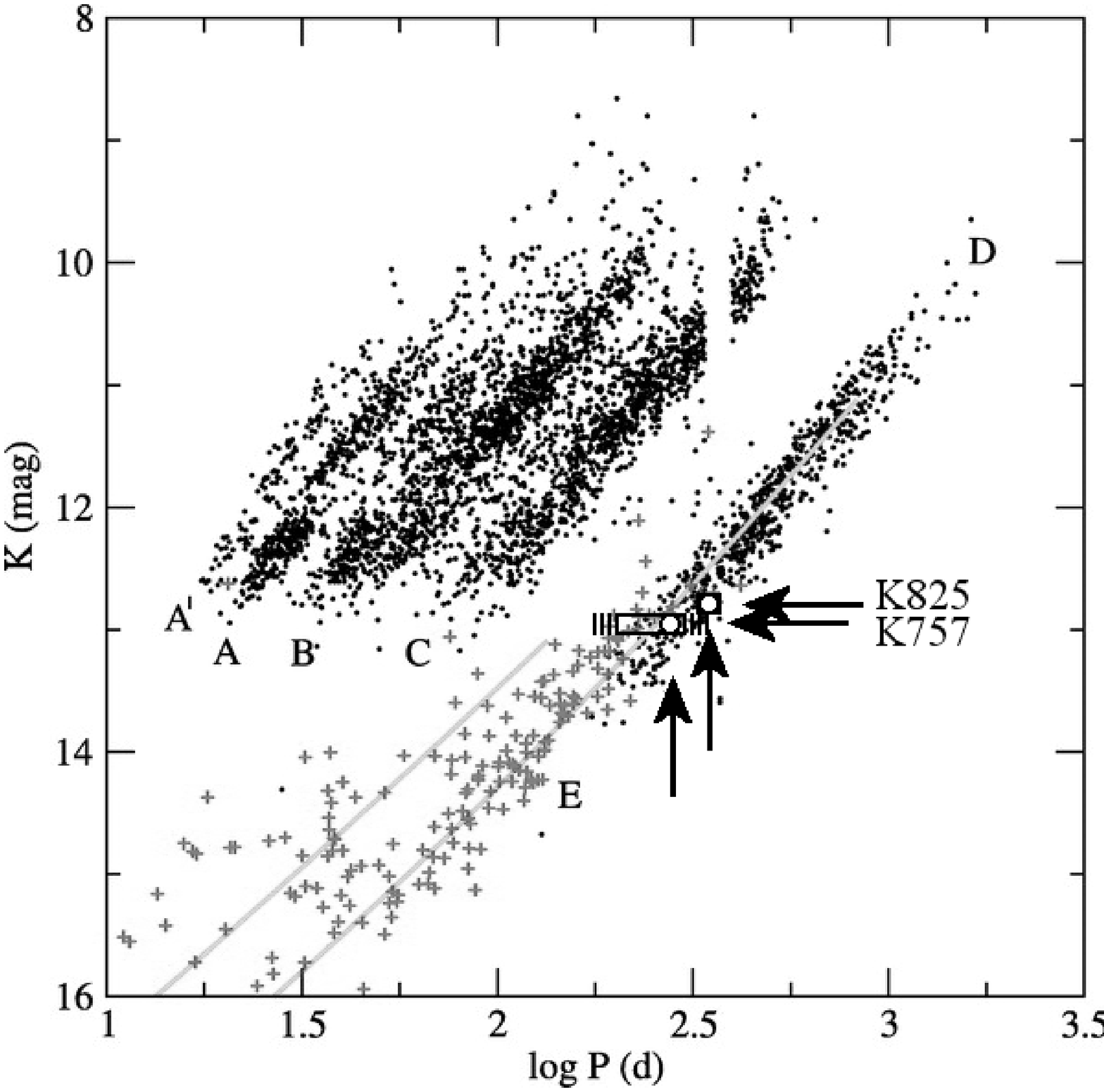}}}
 \caption[Derekas et al.\ Fig.\ 2]{Period--luminosity diagram for the Large Magellanic Cloud (LMC), reproduced from \protect\citet{DKB+06}, their Fig.\ 2. Black dots show pulsating stars, grey plus signs show contact eclipsing binaries. Solid lines show model calculations using evolutionary tracks and Roche geometry. K825 and K757, as they would appear at the distance of the LMC, are shown by the white dots with black borders, highlighted by arrows. Representative uncertainties of these stars' periods are shown. The error on K757's period is poorly defined, but likely falls within the range of the vertical lines. Sequences A$^\prime$, A and B represent pulsation modes, which may be harmonically-related to the fundamental mode, C. Sequence D is the long-secondary period mode, and sequence E is attributed to orbital periods of contact binary stars.}
 \label{M15DKBFig}
\end{figure}

\begin{figure}
\centerline{\resizebox{\hsize}{!}{\includegraphics[angle=0]{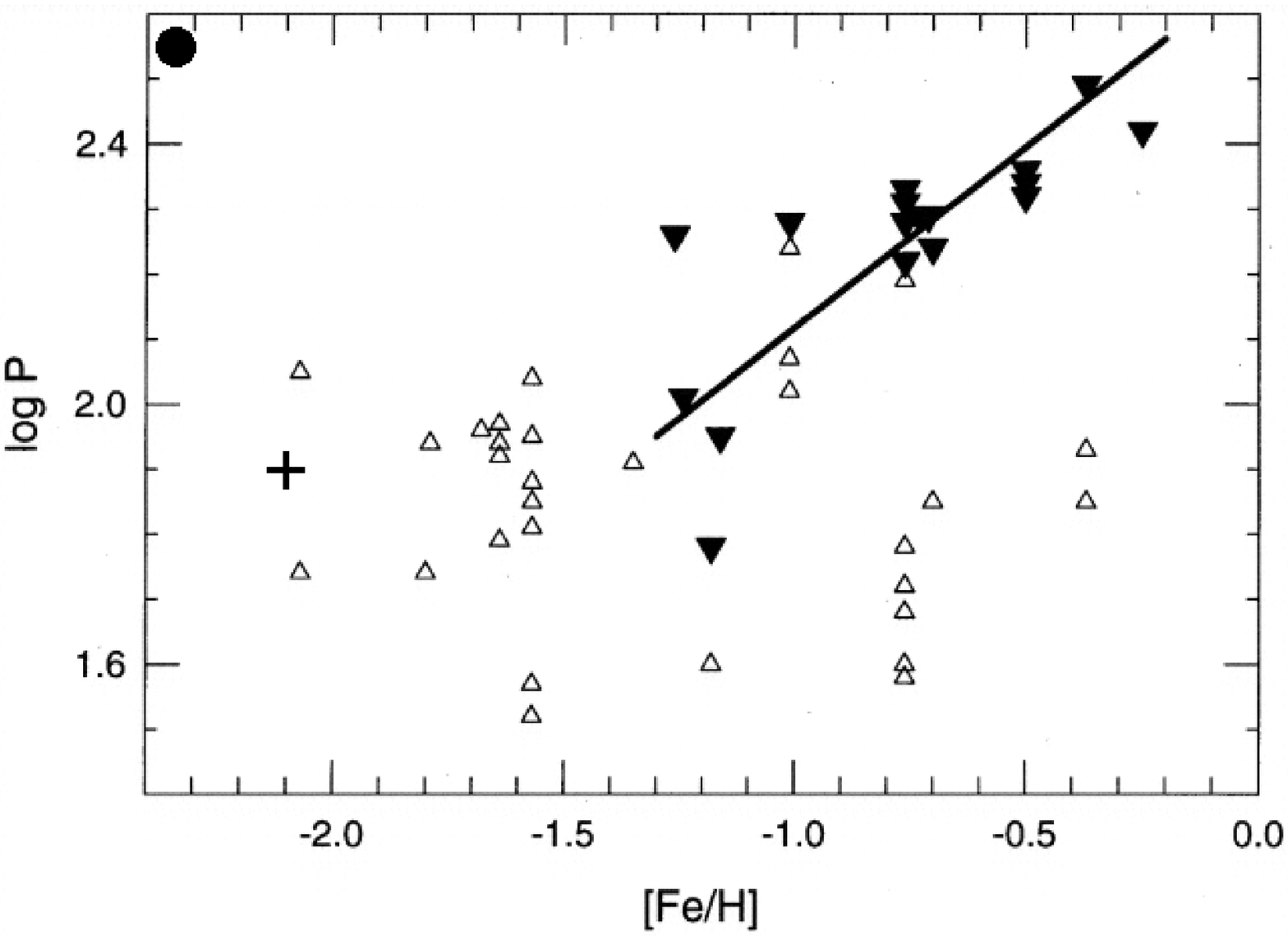}}}
 \caption[Frogel \& Whitelock Fig.\ 4]{Reproduction of Fig.\ 4 from \protect\citet{FW98}, showing luminous LPVs and semi-regular variables (triangles). Filled triangles show the brightest LPV in each cluster (see Frogel \& Whitelock for full selection criteria) and the line denotes their fit. The approximate location of K825 is indicated by the filled circle in the upper-left; NGC 2419 V10 is indicated by a plus sign. The uncertainties for the period and metallicity of K825 are of a similar size to that of its marker.}
 \label{M15FWFig}
\end{figure}

As a populous cluster, we may therefore na\"{i}vely expect M15 to contain a relatively large number of high-amplitude LPVs and SRVs. For comparison, $\omega$ Cen --- which is five times as massive and 0.7 dex more metal rich than M15 --- has $\sim$20 variables with periods $>$20 days and $V$-band amplitudes $>$0.2 mag (\citealt{Clement97}; \citealt{MHB04}; \citealt{PdMPB04}; \citealt{vdVvdBVdZ06}). By this estimate, we might expect 4$\pm$2 LPVs in M15. Finding two LPVs is therefore not unexpected, especially given our incomplete coverage. The absence of any \emph{high amplitude} LPVs in M15 is notable, however. This confirms that stellar pulsation plays less of a r\^ole in stellar evolution at substantially sub-solar metallicities \citep{FW98}, but is still present among the oldest, most-metal-poor giants we see.

The long periods of K757 and K825 are unusual. On a $P-L$ diagram (Fig.\ \ref{M15DKBFig}), they fall neatly onto Wood's sequence D (\citealt{WAA+99}; \citealt{Wood00}). This sequence denotes pulsators with long, \emph{secondary} periods (LSPs), though we find no shorter-period pulsations that would fit sequence C (at $\sim$90 days) or higher harmonics (cf. Fig.\ \ref{M15FFTFig}). However, the two stars also fit sequence E from Derekas et al., which is attributed to ellipsoidal variation in red giant (contact) binaries. Note, however, that K825 shows a more temporally asymmetric variability than many similar stars on sequence E (cf. 2.5873.24 in Derekas et al., Fig.~2). For comparison, the unconfirmed B\"ootes I dwarf galaxy member would lie near sequence C.


Pulsation period is general thought to decrease towards low metallicity among red giants of similar age in Galactic globular clusters \citep{FW98}. Were this the case, it would be extremely surprising to find K757 and K825 placed so neatly on one Wood's sequences. Furthermore, K825 shows a period as long as the star with the longest period in Frogel \& Whitelock's sample (NGC 5927 V3, at 312 days), but at a metallicity where no variables this long are found; the closest analogue in \citet{Clement97} is NGC 2419 V10, a semi-regular variable with $P = 81.3$ days and [Fe/H] = --2.12. Frogel \& Whitelock's data do not include all globular cluster variables, and they only claim their correlation of pulsation period and metallicity exists for [Fe/H] $> -1.3$. It appears that, while pulsation amplitude declines with metallicity, pulsation period does not change appreciably.

In Fig.\ \ref{M15FWFig} we reproduce Frogel \& Whitelock's Fig.\ 4 with K825 and NGC 2419 V10 included. From these data, it appears as though no variables with $P > 200$ days and [Fe/H] $< -1.3$ had been discovered prior to \citet{FW98}. We appear to now have found stars in this missing region. The apparent lower luminosity of the RGB tip and lack of high-luminosity AGB stars in M15 (M15 has no obvious AGB stars (\S\ref{HRDSect}), in contrast with intermediate-metallicity clusters, e.g., \citealt{vLMO+06}; \citealt{BMvL+09}; \citealt{McDonald09}) means that these stars are on the LSP sequence D, rather than appearing in sequence C like their metal-intermediate counterparts.

\subsubsection{Intracluster medium and dust production}

High stellar temperatures and low metallicity make it hard to form circumstellar dust around stars in M15. Despite this, M15 remains the only cluster in which intracluster gas and dust have been detected (\citealt{ESvL+03}; \citealt{vLSEM06}; \citealt{BWvL+06}; \citealt{vLSP+09}). These intracluster media must have been produced in the (astronomically-)recent past ($\sim$1 Myr). However, while K757 and K825 both show evidence for gaseous mass loss (\citealt{MDS08}; \citealt{MAD09}), neither star appears to be producing dust \citep{BWvL+06} suggesting dust production in M15 may not be dominated by stars near and above the RGB tip, as seen in other, more-metal-rich clusters (e.g.\ $\omega$ Centauri, at [Fe/H] $\sim$ --1.6; and 47 Tuc at [Fe/H] = --0.7 --- \citealt{vLMO+06}; \citealt{LPH+06}; \citealt{ORFF+07}; \citealt{MvLD+09}).  

Two likely possibilities remain to explain the presence of intra-cluster dust in M15. Dust production in individual stars could be episodic (e.g.\ \citealt{OFFPR02}, 2007; \citealt{MvLD+09}). Alternatively, relatively-massive ($M_{\rm H} \sim $0.1 M$_\odot$) amounts of dust-containing medium could be produced in a superwind at the AGB (and possibly RGB) tips, or in creating a post-AGB star. Indeed, Boyer et al.\ observe dust in M15's planetary nebula, Pease 1. However, the amount of intra-cluster medium ($M_{\rm H} \sim $0.3 M$_\odot$; \citealt{vLSEM06}) would require this short phase to have recently occurred near-simultaneously (within $\sim$1 Myr) for at least three stars. As a result, we expect that episodic dust production around stars with more-sustained gaseous outflows seems the more-likely explanation for the presence of intra-cluster dust, though both factors could contribute.

If dust production is episodic, would we expect to see dust in the most-evolved stars? Of the 17 most-evolved stars in M15 (Fig.\ \ref{M15LTCMD3}), only K479, K421 and K373 show clear infrared excess in \citet{BWvL+06}. K204 may also show excess, but it is blended with K570 in the \emph{Spitzer} Space Telescope 3--70-$\mu$m photometry. Assuming these stars have luminosities of 1400--1600 L$_\odot$ (based on \S\ref{K825Sect}), we can compare this to Fig.\ 20 of \citet{MvLD+09}, which shows the fraction of dusty stars as a function of luminosity in $\omega$ Cen. Assuming a detection limit for circumstellar dust corresponding approximately to the 6$\sigma$ line in McDonald et al.\ (this approximately takes into account the metallicity difference between the clusters), we might presume that $\sim$30\% (five) of these 17 stars would have circumstellar dust, which is statistically indifferent from the three or four observed.

The lack of dust production in K757 and K825 is puzzling, however, when one considers that the above H$\alpha$ observations (\S\ref{Halpha}; \citealt{MAD09}) indicate gaseous mass loss does occur from these stars, and indeed from stars further down the giant branch (\citealt{MvL07}; \citealt{MAD09}). The pulsation period is long enough for dust to form between pulsations, but not long enough for dust to become so cool as to be undetectable in its [8]--[24] colour (assuming $v_\infty \lesssim 50$ km s$^{-1}$). The pulsations also appear not to generate strong shock waves (which show up as sharp peaks in H$\alpha$ --- e.g.\ \citealt{MvL07}), which may assist dust production. The constant, moderate-velocity outflow provided by the chromosphere may also prevent the formation of a quasi-static molecular layer required to seed dust formation \citep{TOAY97}; the formation of such a layer being additionally hindered by the hotter photospheric temperatures and reduced availability of metals in such low-metallicity stars. Furthermore, it may be that the low amplitude of the pulsations means the scaleheight of the atmosphere is not sufficiently raised, meaning the wind density beyond the dust-formation radius is too low to efficiently produce dust, while the increased effective temperature of the metal-poor stars means the dust-formation radius is too far from the star.

It is clear from observations of more metal-rich clusters (\citealt{RJ01}; \citealt{vLMO+06}; \citealt{LPH+06}; \citealt{MvLD+09}) that efficient dust production occurs mainly in strongly variable stars at the upper end of the RGB, and that the amount of mass loss roughly correlates with the strength of the pulsation as traced by the optical photometric amplitude. Our observations here confirm that correlation, finding no detectable dust production in stars which (while at the RGB tip) exhibit only weak pulsation.

\section{Conclusions}

This study presents a search for long-period variability among giants in the globular cluster M15 and has confirmed the most metal-poor variables known in our Galaxy: K757 and K825. These stars are very close to the RGB-tip and show lightcurves characteristic of pulsation, albeit with very low amplitude. Their periods place them on the `long secondary period' (LSP) sequence (\citealt{WAA+99}; \citealt{Wood00}), though no `primary' period has been found.

The pulsational velocities are similar to those measured at the base of the wind, but it is not certain that pulsation is required to launch or drive the wind. In any case, these pulsations appear not to lead to dust production, which must be caused by a different mechanism, possibly episodically. Although we do find evidence for acceleration in the singly-ionised calcium line profiles, radiation pressure on grains cannot therefore be held responsible for driving the wind. Instread, Alfv\'en waves may couple to a weakly-ionised gas and thus drive the wind, something which could be facilitated at lower metallicity \citep{vLOG+10}.

\section*{Acknowledgments}

I.M.~was supported by a PPARC/STFC studentship for the initial stages of this work. This paper includes data gathered with the 6.5\,m Magellan Telescopes located at Las Campanas Observatory, Chile; the Liverpool Telescope, operated on the island of La Palma by John Moores University in the Spanish Observatorio del Roque de los Muchachos of the Instituto de Astrophysica de Canarias with financial support from the UK Science and Technology Facilities Council; and the Keele Thornton Telescope operated by and located at Keele University, with dedicated support from local volunteers.


\begin{thebibliography}{65}
\expandafter\ifx\csname natexlab\endcsname\relax\def\natexlab#1{#1}\fi

\bibitem[{{Alonso} {et~al.}(2000){Alonso}, {Salaris}, {Arribas},
  {Mart{\'{\i}}nez-Roger}, \& {Asensio Ramos}}]{ASA+00}
{Alonso} A., {Salaris} M., {Arribas} S., {Mart{\'{\i}}nez-Roger} C., {Asensio
  Ramos} A., 2000, A\&A, 355, 1060

\bibitem[{{Arp}(1955)}]{Arp55}
{Arp} H.~C., 1955, AJ, 60, 317

\bibitem[{{Bao-An}(1990)}]{Bao-An90}
{Bao-An} Y., 1990, IBVS, 3431, 1

\bibitem[{{Berry} \& {Burnell}(2005)}]{BB05}
{Berry} R., {Burnell} J., 2005, {The handbook of astronomical image processing}, 2nd ed.\ Richmond,
  VA: Willmann-Bell, 2005

\bibitem[{{Bowen}(1988)}]{Bowen88}
{Bowen} G.~H., 1988, in {Stalio} R., {Willson} L.~A., eds., ASSL Vol. 148: Pulsation and Mass Loss in Stars,
  p.~3


\bibitem[{{Boyer} {et~al.}(2009){Boyer}, {McDonald}, {van Loon}, {Gordon},
  {Babler}, {Block}, {Bracker}, {Engelbracht}, {Hora}, {Indebetouw}, {Meade},
  {Meixner}, {Misselt}, {Oliveira}, {Sewilo}, {Shiao}, \& {Whitney}}]{BMvL+09}
{Boyer} M.~L., {et al.}, 2009, ApJ, 705, 746

\bibitem[{{Boyer} {et~al.}(2006){Boyer}, {Woodward}, {van Loon}, {Gordon},
  {Evans}, {Gehrz}, {Helton}, \& {Polomski}}]{BWvL+06}
{Boyer} M.~L., {Woodward} C.~E., {van Loon} J.~T., {Gordon} K.~D., {Evans} A.,
  {Gehrz} R.~D., {Helton} L.~A., {Polomski} E.~F., 2006, AJ, 132, 1415

\bibitem[{{Brown}(1951)}]{Brown51}
{Brown} A., 1951, ApJ, 113, 344

\bibitem[{{Buonanno} {et~al.}(1983){Buonanno}, {Buscema}, {Corsi}, {Iannicola},
  \& {Fusi Pecci}}]{BBC+83}
{Buonanno} R., {Buscema} G., {Corsi} C.~E., {Iannicola} G., {Fusi Pecci} F.,
  1983, A\&AS, 51, 83

\bibitem[{{Cacciari} {et~al.}(2004){Cacciari}, {Bragaglia}, {Rossetti}, {Fusi
  Pecci}, {Mulas}, {Carretta}, {Gratton}, {Momany}, \& {Pasquini}}]{CBR+04}
{Cacciari} C., {Bragaglia} A., {Rossetti} E., {Fusi Pecci} F., {Mulas} G.,
  {Carretta} E., {Gratton} R.~G., {Momany} Y., {Pasquini} L., 2004, A\&A, 413,
  343

\bibitem[{{Clement}(1997)}]{Clement97}
{Clement} C., 1997, VizieR Online Data Catalog, 5097, 0

\bibitem[{{Cort{\'e}s} {et~al.}(2009){Cort{\'e}s}, {Silva}, {Recio-Blanco},
  {Catelan}, {Do Nascimento}, \& {De Medeiros}}]{CSRB+09}
{Cort{\'e}s} C., {Silva} J.~R.~P., {Recio-Blanco} A., {Catelan} M., {Do
  Nascimento} J.~D., {De Medeiros} J.~R., 2009, ApJ, 704, 750

\bibitem[{{Cudworth}(1976)}]{Cudworth76}
{Cudworth} K.~M., 1976, AJ, 81, 519

\bibitem[{{Dall'Ora} {et~al.}(2006){Dall'Ora}, {Clementini}, {Kinemuchi},
  {Ripepi}, {Marconi}, {di Fabrizio}, {Greco}, {Rodgers}, {K{\"u}hn}, \&
  {Smith}}]{DOCK+06}
{Dall'Ora} M., {Clementini} G., {Kinemuchi} K., {Ripepi} V., {Marconi} M., {di
  Fabrizio} L., {Greco} C., {Rodgers} C.~T., {K{\"u}hn} C., {Smith} H.~A.,
  2006, ApJ, 653, L109

\bibitem[{{Derekas} {et~al.}(2006){Derekas}, {Kiss}, {Bedding}, {Kjeldsen},
  {Lah}, \& {Szab{\'o}}}]{DKB+06}
{Derekas} A., {Kiss} L.~L., {Bedding} T.~R., {Kjeldsen} H., {Lah} P.,
  {Szab{\'o}} G.~M., 2006, ApJ, 650, L55

\bibitem[{{Dickens} {et~al.}(1972){Dickens}, {Feast}, \& {Lloyd
  Evans}}]{DFLE72}
{Dickens} R.~J., {Feast} M.~W., {Lloyd Evans} T., 1972, MNRAS, 159, 337

\bibitem[{{Dupree} {et~al.}(1984){Dupree}, {Hartmann}, \& {Avrett}}]{DHA84}
{Dupree} A.~K., {Hartmann} L., {Avrett} E.~H., 1984, ApJ, 281, L37

\bibitem[{{Dupree} {et~al.}(1999){Dupree}, {Whitney}, \& {Pasquini}}]{DWP99}
{Dupree} A.~K., {Whitney} B.~A., {Pasquini} L., 1999, ApJ, 520, 751

\bibitem[{{Dupree} {et~al.}(2005){Dupree}, {Lobel}, {Young}, {Ake}, {Linsky},
  \& {Redfield}}]{DLY+05}
{Dupree} A.~K., {Lobel} A., {Young} P.~R., {Ake} T.~B., {Linsky} J.~L.,
  {Redfield} S., 2005, ApJ, 622, 629

\bibitem[{{Dupree} {et~al.}(2007){Dupree}, {Li}, \& {Smith}}]{DLS07}
{Dupree} A.~K., {Li} T.~Q., {Smith} G.~H., 2007, AJ, 134, 1348

\bibitem[{{Dupree} {et~al.}(2009){Dupree}, {Smith}, \& {Strader}}]{DSS09}
{Dupree} A.~K., {Smith} G.~H., {Strader} J., 2009, AJ, 138, 1485

\bibitem[{{Evans} {et~al.}(2003){Evans}, {Stickel}, {van Loon}, {Eyres},
  {Hopwood}, \& {Penny}}]{ESvL+03}
{Evans} A., {Stickel} M., {van Loon} J.~T., {Eyres} S.~P.~S., {Hopwood}
  M.~E.~L., {Penny} A.~J., 2003, A\&A, 408, L9

\bibitem[{{Frogel} {et~al.}(1983){Frogel}, {Persson}, \& {Cohen}}]{FPC83}
{Frogel} J.~A., {Persson} S.~E., {Cohen} J.~G., 1983, ApJS, 53, 713

\bibitem[{{Frogel} \& {Whitelock}(1998)}]{FW98}
{Frogel} J.~A., {Whitelock} P.~A., 1998, AJ, 116, 754

\bibitem[{{Gebhardt} {et~al.}(1997){Gebhardt}, {Pryor}, {Williams}, {Hesser},
  \& {Stetson}}]{GPW+97}
{Gebhardt} K., {Pryor} C., {Williams} T.~B., {Hesser} J.~E., {Stetson} P.~B.,
  1997, AJ, 113, 1026

\bibitem[{Harris(1996)}]{Harris96}
Harris W.~E., 1996, ApJ, 112, 1487

\bibitem[{{Kustner}(1921)}]{Kustner21}
{Kustner} F., 1921, Ver\"{o}ffentlichungen des Astronomisches Institute der
  Universit\"{a}t Bonn, 15, 1

\bibitem[{{Lebzelter} {et~al.}(2006){Lebzelter}, {Posch}, {Hinkle}, {Wood}, \&
  {Bouwman}}]{LPH+06}
{Lebzelter} T., {Posch} T., {Hinkle} K., {Wood} P.~R., {Bouwman} J., 2006, ApJ,
  653, L145

\bibitem[{{Lebzelter} {et~al.}(2005){Lebzelter}, {Wood}, {Hinkle}, {Joyce}, \&
  {Fekel}}]{LWH+05}
{Lebzelter} T., {Wood} P.~R., {Hinkle} K.~H., {Joyce} R.~R., {Fekel} F.~C.,
  2005, A\&A, 432, 207

\bibitem[{{Marshall} {et~al.}(2004){Marshall}, {van Loon}, {Matsuura}, {Wood},
  {Zijlstra}, \& {Whitelock}}]{MvLM+04}
{Marshall} J.~R., {van Loon} J.~T., {Matsuura} M., {Wood} P.~R., {Zijlstra}
  A.~A., {Whitelock} P.~A., 2004, MNRAS, 355, 1348

\bibitem[{{Mauas} {et~al.}(2006){Mauas}, {Cacciari}, \& {Pasquini}}]{MCP06}
{Mauas} P.~J.~D., {Cacciari} C., {Pasquini} L., 2006, A\&A, 454, 609

\bibitem[{{McDonald}(2009)}]{McDonald09}
{McDonald} I., 2009, Ph.D.~Thesis, Keele Univ.

\bibitem[{{McDonald} \& {van Loon}(2007)}]{MvL07}
{McDonald} I., {van Loon} J.~T., 2007, A\&A, 476, 1261

\bibitem[{{McDonald} {et~al.}(2009){McDonald}, {van Loon}, {Decin}, {Boyer},
  {Dupree}, {Evans}, {Gehrz}, \& {Woodward}}]{MvLD+09}
{McDonald} I., {van Loon} J.~T., {Decin} L., {Boyer} M.~L., {Dupree} A.~K.,
  {Evans} A., {Gehrz} R.~D., {Woodward} C.~E., 2009, MNRAS, 394, 831

\bibitem[{{McNamara} {et~al.}(2004){McNamara}, {Harrison}, \&
  {Baumgardt}}]{MHB04}
{McNamara} B.~J., {Harrison} T.~E., {Baumgardt} H., 2004, ApJ, 602, 264

\bibitem[{{Menzies} \& {Whitelock}(1985)}]{MW85}
{Menzies} J.~W., {Whitelock} P.~A., 1985, MNRAS, 212, 783

\bibitem[{{M{\'e}sz{\'a}ros} {et~al.}(2008{\natexlab{a}}){M{\'e}sz{\'a}ros},
  {Dupree}, \& {Szentgy{\"o}rgyi}}]{MDS08}
{M{\'e}sz{\'a}ros} S., {Dupree} A.~K., {Szentgy{\"o}rgyi} A.,
  2008{\natexlab{a}}, AJ, 135, 1117

\bibitem[{{M{\'e}sz{\'a}ros} {et~al.}(2008{\natexlab{b}}){M{\'e}sz{\'a}ros},
  {Dupree}, \& {Szentgy{\"o}rgyi}}]{MDS+08}
---, 2008{\natexlab{b}}, AJ, 135, 1117

\bibitem[{{M{\'e}sz{\'a}ros} {et~al.}(2009){M{\'e}sz{\'a}ros}, {Avrett}, \&
  {Dupree}}]{MAD09}
{M{\'e}sz{\'a}ros} S., {Avrett} E.~H., {Dupree} A.~K., 2009, AJ, 138, 615

\bibitem[{{Mosley} \& {White}(1975)}]{MW75}
{Mosley} D.~R., {White} R.~E., 1975, in Bull.~Amer.~Astron.~Soc., Vol.~7,
  p. 535

\bibitem[{{Nicholls} {et~al.}(2009){Nicholls}, {Wood}, {Cioni}, \&
  {Soszy{\'n}ski}}]{NWCS09}
{Nicholls} C.~P., {Wood} P.~R., {Cioni} M., {Soszy{\'n}ski} I., 2009, MNRAS,
  399, 2063

\bibitem[{{Origlia} {et~al.}(2002){Origlia}, {Ferraro}, {Fusi Pecci}, \&
  {Rood}}]{OFFPR02}
{Origlia} L., {Ferraro} F.~R., {Fusi Pecci} F., {Rood} R.~T., 2002, ApJ, 571,
  458

\bibitem[{{Origlia} {et~al.}(2007){Origlia}, {Rood}, {Fabbri}, {Ferraro}, {Fusi
  Pecci}, \& {Rich}}]{ORFF+07}
{Origlia} L., {Rood} R.~T., {Fabbri} S., {Ferraro} F.~R., {Fusi Pecci} F.,
  {Rich} R.~M., 2007, ApJ, 667, L85

\bibitem[{{Pasquali} {et~al.}(2004){Pasquali}, {de Marchi}, {Pulone}, \&
  {Brigas}}]{PdMPB04}
{Pasquali} A., {de Marchi} G., {Pulone} L., {Brigas} M.~S., 2004, A\&A, 428,
  469

\bibitem[{{Pilachowski} {et~al.}(2000){Pilachowski}, {Sneden}, {Kraft},
  {Harmer}, \& {Willmarth}}]{PSK+00}
{Pilachowski} C.~A., {Sneden} C., {Kraft} R.~P., {Harmer} D., {Willmarth} D.,
  2000, AJ, 119, 2895

\bibitem[{{Ramdani} \& {Jorissen}(2001)}]{RJ01}
{Ramdani} A., {Jorissen} A., 2001, A\&A, 372, 85

\bibitem[{{Sandage}(1970)}]{Sandage70}
{Sandage} A., 1970, ApJ, 162, 841

\bibitem[{{Smoker} {et~al.}(2002){Smoker}, {Keenan}, {Lehner}, \&
  {Trundle}}]{SKLT02}
{Smoker} J.~V., {Keenan} F.~P., {Lehner} N., {Trundle} C., 2002, A\&A, 387,
  1057

\bibitem[{{Sneden} {et~al.}(1997){Sneden}, {Kraft}, {Shetrone}, {Smith},
  {Langer}, \& {Prosser}}]{SKS+97}
{Sneden} C., {Kraft} R.~P., {Shetrone} M.~D., {Smith} G.~H., {Langer} G.~E.,
  {Prosser} C.~F., 1997, AJ, 114, 1964

\bibitem[{{Sneden} {et~al.}(2000){Sneden}, {Pilachowski}, \& {Kraft}}]{SPK00}
{Sneden} C., {Pilachowski} C.~A., {Kraft} R.~P., 2000, AJ, 120, 1351

\bibitem[{{Steele} {et~al.}(2004){Steele}, {Smith}, {Rees}, {Baker}, {Bates},
  {Bode}, {Bowman}, {Carter}, {Etherton}, {Ford}, {Fraser}, {Gomboc}, {Lett},
  {Mansfield}, {Marchant}, {Medrano-Cerda}, {Mottram}, {Raback}, {Scott},
  {Tomlinson}, \& {Zamanov}}]{SSR+04}
{Steele} I.~A., {et al.}, 2004, in {J.~M.~Oschmann Jr.}, ed., Society of Photo-Optical
  Instrumentation Engineers (SPIE) Conference Series, Vol. 5489, p. 679

\bibitem[{{Stetson}(1987)}]{Stetson87}
{Stetson} P.~B., 1987, PASP, 99, 191

\bibitem[{{Stetson}(1994)}]{Stetson94}
---, 1994, PASP, 106, 250

\bibitem[{{Tsuji} {et~al.}(1997){Tsuji}, {Ohnaka}, {Aoki}, \&
  {Yamamura}}]{TOAY97}
{Tsuji} T., {Ohnaka} K., {Aoki} W., {Yamamura} I., 1997, A\&A, 320, L1

\bibitem[{{van de Ven} {et~al.}(2006){van de Ven}, {van den Bosch}, {Verolme},
  \& {de Zeeuw}}]{vdVvdBVdZ06}
{van de Ven} G., {van den Bosch} R.~C.~E., {Verolme} E.~K., {de Zeeuw} P.~T.,
  2006, A\&A, 445, 513

\bibitem[{{van Loon}(2000)}]{vanLoon00}
{van Loon} J.~T., 2000, A\&A, 354, 125

\bibitem[{{van Loon} {et~al.}(2006{\natexlab{a}}){van Loon}, {McDonald},
  {Oliveira}, {Evans}, {Boyer}, {Gehrz}, {Polomski}, \& {Woodward}}]{vLMO+06}
{van Loon} J.~T., {McDonald} I., {Oliveira} J.~M., {Evans} A., {Boyer} M.~L.,
  {Gehrz} R.~D., {Polomski} E., {Woodward} C.~E., 2006{\natexlab{a}}, A\&A,
  450, 339

\bibitem[{{van Loon} {et~al.}(2006{\natexlab{b}}){van Loon},
  {Stanimirovi{\'c}}, {Evans}, \& {Muller}}]{vLSEM06}
{van Loon} J.~T., {Stanimirovi{\'c}} S., {Evans} A., {Muller} E.,
  2006{\natexlab{b}}, MNRAS, 365, 1277

\bibitem[{{van Loon} {et~al.}(2007){van Loon}, {van Leeuwen}, {Smalley},
  {Smith}, {Lyons}, {McDonald}, \& {Boyer}}]{vLvLS+07}
{van Loon} J.~T., {van Leeuwen} F., {Smalley} B., {Smith} A.~W., {Lyons} N.~A.,
  {McDonald} I., {Boyer} M.~L., 2007, MNRAS, 382, 1353

\bibitem[{{van Loon} {et~al.}(2009){van Loon}, {Stanimirovi{\'c}}, {Putman},
  {Peek}, {Gibson}, {Douglas}, \& {Korpela}}]{vLSP+09}
{van Loon} J.~T., {Stanimirovi{\'c}} S., {Putman} M.~E., {Peek} J.~E.~G.,
  {Gibson} S.~J., {Douglas} K.~A., {Korpela} E.~J., 2009, MNRAS, 396, 1096

\bibitem[{{van Loon} {et~al.}(2010){van Loon}, {Oliveira}, {Gordon}, {Meixner},
  {Shiao}, {Boyer}, {Kemper}, {Woods}, {Tielens}, {Marengo}, {Indebetouw},
  {Sloan}, \& {Chen}}]{vLOG+10}
{van Loon} J.~T., {et al.}, 2010, AJ, 139, 68

\bibitem[{{Welty}(1985)}]{Welty85}
{Welty} D.~E., 1985, AJ, 90, 2555

\bibitem[{{Wood}(2000)}]{Wood00}
{Wood} P.~R., 2000, 17, 18

\bibitem[{{Wood} {et~al.}(1999){Wood}, {Alcock}, {Allsman}, {Alves}, {Axelrod},
  {Becker}, {Bennett}, {Cook}, {Drake}, {Freeman}, {Griest}, {King}, \& {et
  al.}}]{WAA+99}
{Wood} P.~R., {et al.}, 1999, in {le Bertre} T., {Lebre} A., {Waelkens} C.,
  eds., IAU Symp.~191: Asymptotic Giant Branch Stars, p. 151

\bibitem[{{Yao} {et~al.}(1993){Yao}, {Zhang}, {Qin}, \& {Tong}}]{YZQT93}
{Yao} B.-A., {Zhang} C.-S., {Qin} D., {Tong} J.-H., 1993, Ap\&SS, 210, 163

\end{thebibliography}

\label{lastpage}

\end{document}